\documentclass[aps,twocolumn,superscriptaddress,nofootinbib,longbibliography,notitlepage,floatfix]{revtex4-2}
\UseRawInputEncoding
\usepackage{amsmath}
\usepackage{amssymb}
\usepackage{cancel}
\usepackage{esint}
\usepackage{bbm}

\makeatletter

\providecommand{\tabularnewline}{\\}

\usepackage{graphicx}
\usepackage[section]{placeins}%%put table in its section
\usepackage{tabularx}
\usepackage{amsmath}
\usepackage{amstext}
\usepackage{amssymb}
\usepackage{xfrac}
\usepackage[colorlinks,citecolor=blue]{hyperref}
\usepackage{graphicx}
\usepackage{amsmath}
\usepackage{amstext}
\usepackage{amssymb}
\usepackage{amsfonts}
\usepackage{longtable,booktabs}
\usepackage{hyperref}
\usepackage{url}
\usepackage{subfigure}
\usepackage{dsfont}
\usepackage{booktabs}
\usepackage{amsbsy}
\usepackage{dcolumn}
\usepackage{amsthm}
\usepackage{bm}
\usepackage{esint}
\usepackage{multirow}
\usepackage{hyperref}
\usepackage{cleveref}
\usepackage{mathrsfs}
\usepackage{amsfonts}
\usepackage{amsbsy}
\usepackage{dcolumn}
\usepackage{bm}
\usepackage{multirow}
\usepackage{color}
\usepackage{extarrows}
\usepackage{datetime}
\usepackage{comment}
\usepackage[super]{nth}
\usepackage{tikz}

\hypersetup{
	colorlinks=true,
	linkcolor=blue,
	filecolor=magenta,
	urlcolor=blue,
}

\newcommand{\comments}[1]{}

\usepackage{scalerel}
\usepackage{tikz}
\usetikzlibrary{svg.path}

\definecolor{orcidlogocol}{HTML}{A6CE39}
\tikzset{
	orcidlogo/.pic={
		\fill[orcidlogocol] svg{M256,128c0,70.7-57.3,128-128,128C57.3,256,0,198.7,0,128C0,57.3,57.3,0,128,0C198.7,0,256,57.3,256,128z};
		\fill[white] svg{M86.3,186.2H70.9V79.1h15.4v48.4V186.2z}
		svg{M108.9,79.1h41.6c39.6,0,57,28.3,57,53.6c0,27.5-21.5,53.6-56.8,53.6h-41.8V79.1z M124.3,172.4h24.5c34.9,0,42.9-26.5,42.9-39.7c0-21.5-13.7-39.7-43.7-39.7h-23.7V172.4z}
		svg{M88.7,56.8c0,5.5-4.5,10.1-10.1,10.1c-5.6,0-10.1-4.6-10.1-10.1c0-5.6,4.5-10.1,10.1-10.1C84.2,46.7,88.7,51.3,88.7,56.8z};
	}
}

\newcommand\orcidicon[1]{\href{https://orcid.org/#1}{\mbox{\scalerel*{
				\begin{tikzpicture}[yscale=-1,transform shape]
					\pic{orcidlogo};
				\end{tikzpicture}
			}{|}}}}
\tikzset{bold/.style={color=blue, line width=2pt}}
\tikzset{redop/.style={circle,fill=red}}
\tikzset{blueop/.style={circle,fill=blue}}

\makeatother

\begin{document}
	\title{Non-invertible symmetries and mixed anomalies from equations of motion in (3+1)D twisted $BF$ topological quantum field theories}
	\author{Zhi-Feng Zhang\orcidicon{0009-0005-5267-4108}}\thanks{These authors contributed equally.}
	\affiliation{Max Planck Institute for the Physics of Complex Systems, N\"othnitzer Stra{\ss}e~38, Dresden~01187, Germany}
	\author{Yizhou Huang\orcidicon{0009-0008-4080-8174}}\thanks{These authors contributed equally.}
	\affiliation{School of Physics, Sun Yat-sen University, Guangzhou, 510275, China}
	\author{Qing-Rui Wang\orcidicon{0000-0002-4850-8511}}
	\email{wangqr@mail.tsinghua.edu.cn}
	\affiliation{Yau Mathematical Sciences Center, Tsinghua University, Haidian, Beijing, China}
	\author{Peng Ye\orcidicon{0000-0002-6251-677X}}
	\email{yepeng5@mail.sysu.edu.cn}
	\affiliation{School of Physics, Sun Yat-sen University, Guangzhou, 510275, China}

	\begin{abstract} 
		We develop a systematic study on the topological operators that generate invertible and non-invertible higher-form symmetries in $\left(3+1\right)D$ twisted $BF$ topological quantum field theories (TQFTs). We focus on the twisted $BF$ theories with gauge group $G=\prod_i \mathbb{Z}_{N_i}$ and an $a\wedge a\wedge b$ twist, where $a$'s and $b$ are 1-form and 2-form gauge fields, respectively. These theories realize non-Abelian $(3+1)$D TQFTs supporting Borromean-Rings braiding and describe three-dimensional non-Abelian topological orders in condensed matter. For $G=(\mathbb{Z}_2)^3$, a microscopic realization is given by the $\mathbb{D}_4$ Kitaev quantum double model. Starting from the equations of motion (EoMs), we construct gauge invariant operators that are supported on closed submanifolds, hence are genuine topological operators. Some of them are associated with nontrivial Hilbert spaces which are the projective representation spaces of specific gauge subgroups, thus 
        being non-invertible. The others are invertible topological operators with group-like fusion rules. The fusion rules of non-invertible topological operators correspond to the decomposition of the product of the projective representation spaces. We provide a field-theoretical algorithm to calculate the fusion rules. Besides, we identify an obstruction to retaining a non-invertible topological operator after gauging an invertible symmetry, and derive the constraints on pairs of non-invertible operators relevant to simultaneous gauging. Our results give a unified and practical approach for identifying and characterizing higher-form symmetries and mixed anomalies, which can be extended to more general TQFTs and topological orders.
	\end{abstract}
	\maketitle
	\tableofcontents

	\section{Introduction\label{sec_introduction}}
	
	Symmetry has long been a central organizing principle in physics. In recent years, the notion of symmetry has been substantially generalized from
	multiple perspectives across condensed matter, high-energy, and mathematical
	physics~\cite{gaiotto_generalized_2015, wen_emergent_2019, mcgreevy_generalized_2023,
		cordova_snowmass_2022, schafer-nameki_ictp_2023, bhardwaj_lectures_2023,
		ji_categorical_2020, luo_lecture_2024}. Two complementary aspects are
	particularly important.
	First, symmetry operations need not act on the entire system; instead, they may
	be supported on submanifolds~\cite{gaiotto_generalized_2015}. As a result,
	symmetry operators can be extended (line/surface) operators, and charged
	operators become extended rather than point-like, leading to the concept of
	higher-form symmetry. Within this framework, the $\mathbb{Z}_2$ symmetry of the
	one-dimensional\footnote{In this paper, ``$(n+1)$D'' refers to
		$(n+1)$-dimensional spacetime with $n$-dimensional real space. We avoid using
		``$n$D'' or ``$(n)$D'' unless otherwise specified. When referring specifically
		to spatial dimensions or to the dimensions of geometric objects, we use ``$n$d'',
		e.g., $3$d ground state, $3$d topological order, and $2$d square lattice.} ($1$d) transverse-field Ising model is an example of a $0$-form
	symmetry.
	Second, the algebraic structure organizing symmetry operators need not be a group~\cite{FROHLICH2007354,Runkel_2011_defects,bhardwaj2018finite,chang_topological_2019,Thorngren_fusion,Thorngren_2013_higher_symmetry,kong_algebraic_2020,ji_categorical_2020,Tiwari_2023_higher_categorical,shao_whats_2024}. More generally, symmetry is
	encoded by the fusion of topological operators/defects, and is naturally
	formulated using the language of category theory, giving rise to non-invertible
	or categorical symmetries.
	
	As an illustration of the first aspect, the $2$d toric code model~\cite{Kitaev2003faulttolerant}
	on a square lattice admits symmetry operators supported on closed loops rather
	than on the full lattice. In condensed-matter language, these loop operators
	commute with the Hamiltonian and act within the topological ground-state sector;
	in the continuum/topological quantum field theory (TQFT) language, they are topological line operators (Wilson
	loops) generating higher-form symmetries. Operators supported on one-dimensional
	paths that intersect these loops transform nontrivially under the symmetry and
	are therefore charged. Such symmetries are referred to as $1$-form symmetries,
	reflecting the co-dimensionality of their support in $2$d real space. At long
	wavelengths, higher-form symmetry operators are topological in the sense that
	their supports are manifold-like and can be continuously deformed without
	affecting their action.
	There exists another important generalization, namely subsystem symmetries,
	whose supports are rigid rather than deformable and are typically fixed to
	specific geometric patterns such as lines, fractals, or more exotic structures~\cite{Haah_code_2011,Vijay_fracton_2015,Vijay_Haah_fracton_2016,Nandkishore_fracton_review,Pretko_fracton_review,Shirley_foliated_2019,LiYe2020PRB,LiYeFracton2021,you2019, You2018, Stephen2022,Yuan_fracton_SF_2019,Chen_fracton_SF_2010,Wang_fracton_SF_2024,Wang_KT_2026,PRXQuantum.5.030342}. In this
	work, however, we focus exclusively on higher-form symmetries and leave
	subsystem symmetries for future investigation.
	
	As an illustration of the second aspect, a representative example of a
	non-invertible symmetry arises in the $1$d transverse-field Ising model at
	criticality, which admits a symmetry operation implementing Kramers--Wannier
	duality. This operation exchanges spin and domain-wall variables and commutes
	with the Hamiltonian, yet it does not admit an inverse in the conventional
	sense: acting with it projects out a subset of states. Such an operation
	therefore realizes a non-invertible symmetry. For non-invertible symmetries,
	composing two symmetry operations may not yield a single operator but rather a
	linear combination of symmetry operators governed by fusion rules. By contrast,
	for invertible symmetries the fusion rules reduce to the familiar group
	composition law. Appendix~\ref{appendix_symm_lattice} introduces the generalized symmetries in lattice models.

Building on the concept of generalized symmetry, a framework known as the symmetry topological field theory (SymTFT) has been rapidly developed~\cite{kaidi2023defect}. Consider a physical quantum system that lives in $\left(d-1+1\right)$D spacetime. The SymTFT states that the symmetry structure of this system can be described by a $\left(d+1\right)$D TQFT, with the physical system realized as a boundary condition of the bulk theory (the ``sandwich'' construction). A symmetry operator in the $\left(d-1+1\right)$D system is formed by taking a topological operator from the $\left(d+1\right)$D TQFT and pushing it onto the physical boundary. Such topological operators reflect the generalized symmetries and generalized charges encoded by the $\left(d+1\right)$D TQFT~\cite{bhardwaj_generalized_2023}. A 't Hooft anomaly is an obstruction to gauging a symmetry. A nontrivial SymTFT does not necessarily imply such an anomaly. Different topological boundary conditions correspond to different global variants (absolute theories) of the same local dynamics, and this makes the SymTFT a natural language for gauging and for anomalies of non-invertible symmetries~\cite{kaidi2023symmetry,cordova_anomalies_2024}. This framework has found broad applications, including the classification of gapped and gapless phases, the diagnosis of generalized-symmetry anomalies, the analysis of renormalization-group flows, the study of mixed-state phases and phase transitions, and so on~\cite{kong_algebraic_2020, ji2021unified,vanBeest:2022fss,kaidi2023defect,Chatterjee_holographic_2023,Huang_holography_2025,Lichtman_bulk_anyon_2020,Moradi_holography_2022,Qi_symmetry_taco_2025,Chen_generalized_Landau_2025,chatterjee2023symmetry, Wen:2024udn, wen2025classification, wen2023bulk, Jia_2025_Cat_continuous_symm, Wang_2025_SymTFT_flavor_symm, luo_topological_2025,Hu_fracton_holography_2026}. To date, SymTFT has been particularly successful for $(2+1)$D bulks and $(1+1)$D boundaries. Extending this framework to $(2+1)$D boundary systems therefore naturally motivates a detailed understanding of generalized symmetries realized in $(3+1)$D TQFTs/topologically ordered phases.

	As effective field theories of topological order, TQFTs are expected to encode
	generalized symmetries in a natural and systematic manner. For instance,
	Wilson operators in Chern--Simons theory~\cite{blok_effective_1990,
		PhysRevB.46.2290, Witten1989} act as generators of $1$-form symmetries.
	Analogous to the role of Chern--Simons theory in describing $2$d topological
	orders, $BF$ theory~\cite{horowitz_quantum_1990, hansson_superconductors_2004},
	together with its twisted variants~\cite{ye16a, Moy_Fradkin2023, YW13a,
		2016arXiv161209298P, PhysRevB.99.235137, YeGu2015, ypdw, yp18prl,
		zhang2021compatible, bti2, bti6}, provides a unified field-theoretical
	framework for characterizing $3$d topological orders and gauged
	symmetry-protected topological phases with particle and loop excitations.
	Geometrically, the canonical $BF$ term $b\,da$ (with $b$ a $2$-form field and
	$a$ a $1$-form field) captures the characteristic particle--loop braiding in
	$3$d topological order.\footnote{Throughout this paper, we adopt a condensed
		notation for differential forms and omit the wedge symbol $\wedge$ when no
		confusion arises. For example, $b\,da$ stands for $b \wedge da$, and similarly
		$a a da$ denotes topological terms of $a \wedge a \wedge da$ type.}
	
	More intricate braiding processes, including multi-loop and Borromean-Rings
	braiding, are encoded by twisted terms such as $a a da$, $a a a a$, and
	$a a b$~\cite{hansson_superconductors_2004, 2016arXiv161209298P, yp18prl,
		PRESKILL199050, PhysRevLett.62.1071, PhysRevLett.62.1221, ALFORD1992251,
		wang_levin1, PhysRevLett.114.031601, string4, jian_qi_14, string5,
		chen_bulk-boundary_2016, Tiwari:2016aa, corbodism3, wan_twisted_2015,
		PhysRevX.6.021015}. These braiding processes are subject to compatibility
	conditions imposed by gauge invariance~\cite{zhang2021compatible}. Additional
	terms, such as the twisted $bb$ term and the $\theta$ term $da\,da$, encode
	emergent fermionic statistics~\cite{PhysRevB.99.235137, bti2,
		Kapustin:2014gua, zhang_continuum_2023} and topological
	responses~\cite{ye16a, YW13a, bti6, TI6, lapa17, Ye:2017aa, witten2016fermion,
		PhysRevB.99.205120}, respectively. Further developments revealed symmetry
	fractionalization on loop excitations, leading to mixed three-loop braiding and
	the classification of symmetry-enriched topological phases in higher
	dimensions~\cite{Ning2018prb, ye16_set, 2016arXiv161008645Y}. The fusion and
	shrinking rules of topological excitations in $3$d topological orders have also
	been investigated within the $BF$-theory framework~\cite{zhang_non-abelian_2023}.
	Motivated by this field-theoretical foundation, a diagrammatic representation
	of higher-dimensional topological orders, including $4$d cases~\cite{Zhang:2021ycl,huang2023fusion},
	was recently developed  in Ref.~\cite{huang2025diagrammatics}. Given this
	progress, a natural question arises: how can generalized symmetries of $3$d
	topological orders---equivalently, of their $(3+1)$D effective TQFT
	descriptions---be systematically identified \emph{directly} from continuum
	field theory, together with their fusion and anomaly data?

	In this work we provide a fully constructive, model-resolved derivation of
	generalized symmetries in a nontrivial family of $(3+1)$D TQFTs/topological
	orders\footnote{Reader's guide (condensed-matter vs.\ TQFT language).
		To minimize translation cost across communities, we will freely use both
		terminologies. In particular, ``$3$d topological order'' refers to the gapped
		topological sector captured by a $(3+1)$D TQFT; point/loop excitations in the
		many-body system correspond to line/surface operators (topological defects) in
		the TQFT; and generalized symmetries are implemented by topological operators supported on submanifolds, whose fusion encodes the symmetry algebra.}. Our approach is based on the gauge invariance required by the theory itself and the topological invariance of operators implied by equations of motion (EoMs). We construct topological line and surface operators that generate invertible as well as non-invertible symmetries. Our construction allows us to systematically analyze the algebraic structure of the topological operators including their fusion rules and mixed anomalies. 
        
	Concretely, we study twisted $BF$ theories with gauge group $G=\prod_i \mathbb{Z}_{N_i}$ and an $a \wedge a \wedge b$ twist~\cite{yp18prl,zhang2021compatible,zhang_non-abelian_2023}, which realize non-Abelian $(3+1)$D TQFTs supporting Borromean-Rings braiding and serve as effective continuum descriptions of certain three-dimensional non-Abelian topological orders. For the special case $G=(\mathbb{Z}_2)^3$, the corresponding non-Abelian topological order admits a microscopic realization in the $\mathbb{D}_4$ Kitaev quantum double lattice model~\cite{huang2025bridge}. Without the $a \wedge a \wedge b$ twist, this theory describes an Abelian topological order where all higher-form symmetries are invertible. This setting provides a concrete arena in which higher-form generalized symmetries, including \emph{intrinsically} non-invertible ones, can be identified and analyzed systematically. We obtain main results as summarized below: (1) The EoMs in the $a\wedge a\wedge b$ twisted $BF$ theories naturally separate into two distinct types. The type-I EoMs simply show that the exterior derivative (curvature) of a gauge field vanishes. They lead to conventional Wilson-type topological operators that generate invertible symmetries. The type-II EoMs indicate that the exterior derivatives of some gauge fields are not vanishing, implying that the Wilson-type operators are not gauge invariant. We introduce auxiliary systems to recover the gauge invariance and hence obtain genuine topological line/surface operators. The auxiliary systems transform projectively under the gauge group. Therefore, the topological operator is associated with a nontrivial projective representation space which is also the Hilbert space of the auxiliary system. It is the nontrivial projective representation space that makes the topological operator non-invertible. (2) The fusion of two topological operators is thus the product of the two projective representation spaces, whose decomposition can be derived based on the representation theory of group theory. We also provide a detailed field-theoretical analysis for the fusion rules. (3) We identify an obstruction to retaining an original non-invertible symmetry operator after gauging an invertible symmetry. For pairs of fundamental non-invertible operators, we derive nontrivial constraints and distinguish these selection rules from a complete proof of an obstruction to simultaneous gauging.
	
	This paper is organized as follows. In Sec.~\ref{sec_review_generalized_symmetry},
	we briefly explain our approach to constructing topological operators, emphasizing the relation between generalized symmetries and topological operators. In Sec.~\ref{sec_symmetry_operator_aab}, we systematically construct the gauge invariant topological operators in the class of $a\wedge a \wedge b$ twisted $BF$ theories, which generate the higher-form invertible/non-invertible symmetries of the theories. Sec.~\ref{sec_fusion_rule} is devoted to the
	analysis of fusion rules of topological operators. In Sec.~\ref{sec_anomaly}, we discuss the mixed anomalies among the higher-form symmetries in the $a\wedge a \wedge b$ twisted $BF$ theory. We conclude in Sec.~\ref{sec_discussion} with a summary
	and an outlook. The technical details are collected in the Appendices.

	\section{Symmetry operators, topological invariance, and equations of motion\label{sec_review_generalized_symmetry}}

    In this section, we sketch our strategy of constructing symmetry operators
in this paper. First, we review that the symmetry operator of a continuous
global ($0$-form) symmetry originates from a conserved current. Naturally,
this correspondence can be generalized to continuous higher-form symmetry,
for which higher-form conserved currents are present. However, in
this paper, we focus on gauge theories whose gauge group is discrete,
even though the theories themselves are formulated by continuous gauge
fields. The higher-form symmetries studied here are discrete and are described
by topological operators. 

A higher-form symmetry is
generated by a topological operator supported on a submanifold, i.e., the symmetry operator is a topological operator\footnote{We will use the terms, symmetry operator, topological operator, and topological  defect, interchangeably in this paper.}. By
topological, we mean that the symmetry action on a charged object
is unchanged if the submanifold is deformed smoothly without crossing
any charged objects. In other words, only the topological class of
the submanifold matters, instead of its shape. For continuous higher-form
symmetries, such topological invariance is guaranteed by their conserved
currents: Stokes' theorem keeps the integral of a conserved current invariant
under a smooth deformation of the integration submanifold. For the gauge
theories with discrete gauge groups (e.g., $\mathbb{Z}_{N}$ $BF$ gauge theory), the EoMs  play the same role:  they also imply the topological invariance. Therefore, we
can construct gauge invariant topological operators based on the EoMs,
which are the desired higher-form symmetry operators. In the untwisted examples and their anomaly-inflow discussion, we use \emph{current analogue} for the Hodge dual of a flat gauge potential to illustrate the formal relation between an EoM and a continuity equation. The gauge potential is gauge dependent, and its integral can shift under large gauge transformations; the gauge invariant observables are the appropriately quantized exponentials. In the twisted theory, we construct the topological operators directly from EoMs and gauge invariance, as developed in Sec.~\ref{sec_symmetry_operator_aab}.  

In the following, we will go through this line of
thinking by examples: global $U\left(1\right)$ symmetry, continuous higher-form symmetries in $\left(3+1\right)$D Maxwell theory, and discrete higher-form symmetries in $\mathbb{Z}_{2}$ $BF$ gauge theory. In relativistic field theory, a continuous symmetry is reflected in a conserved current, and the corresponding conserved charge exponentiates to the symmetry operator. Let us consider the global $U\left(1\right)$ symmetry corresponding to particle-number conservation in $(3+1)$D. The spacetime current of particles satisfies the conservation law $\partial_{\mu} j_{\mu}=0$, where $j_{0}$ denotes the particle-number density and $j_{i}$ denotes the particle-current density. This continuity equation can be written as $d\left(* j\right)=0$, where $*$ denotes the Hodge dual. The total particle number of the $3$d system is given by $Q=\int_{\Sigma_{3}}\rho\left(\vec{x},t\right)dx^{1}dx^{2}dx^{3}
	=\int_{\Sigma_{3}}* j$,
	where $\Sigma_{3}$ is the $3$d space at a time slice. Let us consider another time slice $\Sigma_{3}'$ such that $\Sigma_{3}'-\Sigma_{3}=\partial M_{4}$ with $M_4$ a four-dimensional manifold. The increase of total particle number is given by
	$\int_{\Sigma_{3}'}* j-\int_{\Sigma_{3}}* j=\int_{\partial M_{4}}* j$ which can be further written as
	$\int_{M_{4}}d\left(* j\right)$ by Stokes' theorem and vanishes due to $d\left(* j\right)=0$.
	Therefore, the particle number $Q$ is independent of the choice of time slice and is a conserved quantity, implying the existence of a symmetry. This conserved particle number generates the global $U\left(1\right)$ symmetry, whose symmetry operator is
	\begin{equation}
		U_\alpha\left(\Sigma_{3}\right)
		=\exp\left({\rm i}\alpha Q\right)
		=\exp\left({\rm i}\alpha\int_{\Sigma_{3}} * j\right)\,,
	\end{equation}
	where the real parameter $\alpha$ takes value in $[0, 2\pi)$ as a compact phase angle. The compactness of $\alpha$ arises from the quantization of $Q$. The existence of this global symmetry can thus be traced back to the presence of a conserved current. The conservation law ensures that $U_\alpha$ is invariant under smooth deformations of $\Sigma_{3}$, i.e., it is a topological operator acting on the entire quantum system.
	
	A continuous higher-form symmetry, e.g., a $p$-form symmetry, generalizes the above structure by allowing the conserved current to be a $(p+1)$-form. Correspondingly, the associated symmetry operator is supported on a $\left(D-1-p\right)$-dimensional submanifold in $(D-1+1)$D spacetime.
	As an illustrative example of higher-form symmetries, we consider the $(3+1)$D Maxwell electromagnetic theory with action
	$S=\frac{1}{4}\int d^{4}x\, F_{\mu\nu}F^{\mu\nu}$,
	where $F_{\mu\nu}=\partial_{\mu}A_{\nu}-\partial_{\nu}A_{\mu}$; for simplicity we set the gauge coupling $e=1$ (equivalently, absorb it into the normalization of fields). See also the similar discussion in, e.g., Refs~\cite{luo_lecture_2024,gaiotto_generalized_2015,mcgreevy_generalized_2023,bhardwaj_lectures_2023}. First, there exists a conserved $2$-form current satisfying $d\left(* J_{m}\right)=0$, with
	$J_{m}^{\mu\nu}=\frac{1}{4\pi}\epsilon^{\mu\nu\lambda\sigma}F_{\lambda\sigma}
	=\frac{1}{2\pi}\left(d\widetilde{A}\right)^{\mu\nu}$. Here $\widetilde{A}$ is defined by $d\widetilde{A}=* dA$. This continuity equation is equivalent to $dF=0$, namely the Bianchi identity. The conserved current $J_{m}$ gives rise to a ``magnetic'' $1$-form symmetry with symmetry operator
	\begin{equation}
		U_{\alpha}^{\left(m\right)}\left(\Sigma_{2}\right)
		=\exp\left({\rm i}\alpha\int_{\Sigma_{2}}* J_{m}\right)
		=\exp\left({\rm i}\frac{\alpha}{2\pi}\int_{\Sigma_{2}}F\right).
	\end{equation}
	The integral $\frac{1}{2\pi}\int_{\Sigma_{2}}F$ counts the number of magnetic monopoles enclosed by the $2$d submanifold $\Sigma_{2}$.
	Meanwhile, there exists another conserved $2$-form current, $J_{e}=F$, satisfying the continuity equation $d\left(* J_{e}\right)=d\left(* F\right)=0$ that holds wherever the electric charges are absent. The conserved current $J_{e}$ generates an ``electric'' $1$-form symmetry with symmetry operator
	\begin{equation}
		U_{\alpha}^{\left(e\right)}\left(\Sigma_{2}\right)
		=\exp\left({\rm i}\alpha\int_{\Sigma_{2}}* J_{e}\right)
		=\exp\left({\rm i}\alpha\int_{\Sigma_{2}}* F\right).
	\end{equation}

    In this paper, we focus on gauge theories with discrete gauge groups. A prototypical example is the $\mathbb{Z}_{2}$ mutual Chern-Simons
theory, which serves as the effective field theory (i.e., a TQFT description)
of a $\mathbb{Z}_{2}$ $2$d topological order,
	\begin{equation}
		S=\int_{M_{3}}\frac{2}{2\pi}a^{1} da^{2}.\label{cs21d}
	\end{equation}
	The lattice realization of this field theory is the celebrated toric code model in $2$d square lattice. Here $a^{1}$ and $a^{2}$ are $1$-form gauge fields. The EoMs are
	$\frac{1}{\pi}da^{1}=0$ and $\frac{1}{\pi}da^{2}=0$. These EoMs state that the gauge fields are flat.
Therefore, one can introduce two quantities $Q_{1}=\frac{1}{\pi}\int_{\gamma}a^{1}$
and $Q_{2}=\frac{1}{\pi}\int_{\gamma}a^{2}$ that are invariant under a smooth
deformation of $\gamma$, where $\gamma$ is a closed loop. These
quantities further give rise to two gauge invariant topological operators, i.e., two symmetry operators,
	\begin{equation}
		U_{e}\left(\gamma\right)
		=\exp\left({\rm i}\int_{\gamma}a^{2}\right)\,,\,\,\,
		U_{m}\left(\gamma\right)
		=\exp\left({\rm i}\int_{\gamma}a^{1}\right)
	\end{equation}
	which satisfy the commutation algebra
	$
	U_{e}\left(\gamma_{1}\right)
	U_{m}\left(\gamma_{2}\right)
	=
	\left(-1\right)^{{\rm Int}\left(\gamma_{1},\gamma_{2}\right)}
	U_{m}\left(\gamma_{2}\right)
	U_{e}\left(\gamma_{1}\right)$.
	Here, ${\rm Int}\left(\gamma_{1},\gamma_{2}\right)$ denotes the intersection number of the two loops on the same $2$d spatial slice. For a physical picture of these topological operators, we can interpret $Q_{2}$ and $Q_{1}$ as integrals of the Hodge duals of the current analogues, $\mathcal{J}_{e} = * \frac{1}{\pi}a^{2}$ and $\mathcal{J}_{m} = * \frac{1}{\pi}a^{1} $, and the EoMs imply $d\left(* \mathcal{J}_{e}\right)=0$ and $d\left(* \mathcal{J}_{m}\right)=0$. Alternatively, we can think that the conserved objects are the holonomy angles $\int_{\gamma} a^{1}$ and $\int_{\gamma} a^{2}$, with the flatness from EoMs as their conservation law.
    
    The compact gauge-field path integral further imposes
	$\int_{\gamma}2 a^{1}=0\mod 2\pi$ and $\int_{\gamma}2 a^{2}=0\mod 2\pi$.
	Consequently, applying $U_{e}\left(\gamma\right)$ or $U_{m}\left(\gamma\right)$ twice yields the identity operation, and both symmetry operators have a $\mathbb{Z}_{2}$ group structure. In this sense, the field theory~(\ref{cs21d}) exhibits a $\mathbb{Z}_{2,e}^{\left(1\right)}\times\mathbb{Z}_{2,m}^{\left(1\right)}$ $1$-form symmetry implemented by $U_{e}\left(\gamma\right)$ and $U_{m}\left(\gamma\right)$, where the superscript ``$\left(1\right)$'' indicates the $1$-form nature of the symmetry and the subscripts label the two symmetry generators.

    In this paper, we focus on generalized symmetries in field theories.

	\section{Field-theoretical construction of symmetry operators\label{sec_symmetry_operator_aab}}
	
	In this section, we construct symmetry operators (topological operators) from EoMs in $(3+1)$D TQFTs that serve as effective field-theoretical descriptions of $3$d topological orders. We begin with the $(3+1)$D $\mathbb{Z}_{N}$ $BF$ theory as a warm-up, which describes $\mathbb{Z}_N$ topological orders. We then turn to the main goal of this paper: identifying symmetry operators in the effective field theory of a class of $3$d non-Abelian topological orders, namely Borromean-Rings (BR) topological orders. According to the discussion in Sec.~\ref{sec_review_generalized_symmetry}, the symmetry operators are nothing else but gauge invariant topological operators. In particular, we show how the principle of gauge invariance and the topological invariance from EoMs can be exploited to obtain intrinsically non-invertible symmetry operators.
	
	A system with BR topological order~\cite{yp18prl,zhang2021compatible,zhang_non-abelian_2023,huang2025bridge} exhibits nontrivial BR braiding statistics. A BR braiding process involves one particle and two loop excitations: the particle is moved around the two loops such that its spatial trajectory and the two loops form Borromean-Rings. Borromean-Rings are three closed curves in $3$d space that are not pairwise linked, yet cannot be separated as a whole. Consequently, there is no net particle-loop braiding phase, since the particle trajectory is not linked with either loop individually. BR braiding is also distinct from multi-loop braiding because it necessarily involves a particle excitation. The corresponding TQFT is a twisted $BF$ theory with an $aab$-type term; see Eq.~(\ref{eq_aab_action}) below. Remarkably, BR topological order supports non-Abelian particle and loop excitations even when all excitations carry gauge charges and fluxes of an Abelian gauge group, making it an ideal playground for exploring generalized symmetries---including non-invertible ones---in $(3+1)$D topological phases/TQFTs.

In the following, we will derive the EoMs of gauge fields, from which we can construct topological operators based on the principle of gauge invariance. For untwisted gauge theory, this procedure is straightforward, as shown in Sec.~\ref{subsec_review_higher-form_BF_theory}. For the BR topological order whose TQFT is a twisted gauge theory, we find two types of EoMs, denoted as type-I
and type-II, respectively, as shown in Sec.~\ref{subsec_normal_type_EoM} and Sec.~\ref{subsec_special_type_EoM}. The type-I EoMs indicate that specific gauge fields are closed, thus one can write down the conventional Wilson-type operators as the topological operators. On the other hand, the type-II EoMs show that the exterior derivatives of some gauge fields are non-vanishing, so the exponentiation of an integral of a gauge field on a closed submanifold does not give a gauge invariant topological operator. An auxiliary system living on the closed submanifold is required to make the total system gauge invariant. Such an auxiliary system can have a nontrivial Hilbert space that is responsible for the non-invertibility of the topological operators. Similar constructions of non-invertible topological operators are also studied in the literature, e.g., Refs.~\cite{kaidi_kramers-wannier-like_2022,Shao_non-inv_SM-2022}.

	\subsection{Higher-form symmetries in $\left(3+1\right)$D $\mathbb{Z}_{N}$ $BF$ theory \label{subsec_review_higher-form_BF_theory}}
    
    The $(3+1)$D $\mathbb{Z}_{N}$ $BF$ theory, as the effective theory of $3$d $\mathbb{Z}_{N}$ topological order, is 
	\begin{equation}
		S=\int_{M_{4}}\frac{N}{2\pi}bda.\label{eqn_bftermN}
	\end{equation}
	Here $b$ and $a$ are $2$-form and $1$-form gauge fields, respectively. The EoMs are
	\begin{equation}
		\frac{N}{2\pi}da=0\quad\text{and}\quad\frac{N}{2\pi}db=0.
    \label{eom_3d_BF}
	\end{equation}
    They lead to the following rescaled holonomy variables 
    \begin{equation}
    Q_{e}\left(\gamma\right)=\frac{N}{2\pi}\int_{\gamma} a\,,\,\,\,Q_{m}\left(\sigma\right)=\frac{N}{2\pi}\int_{\sigma}b,
	\end{equation}
	where $\gamma$ and $\sigma$ are closed $1$d and $2$d submanifolds, respectively. These variables are defined modulo $N$, since large gauge transformations shift the underlying integrals by multiples of $2\pi$. Their values modulo $N$ are invariant under smooth deformations of $\gamma$ or $\sigma$ that avoid other insertions. To see this, consider a smooth deformation $\gamma_1\rightarrow\gamma_2$ and a surface $\Sigma$ with $\partial\Sigma=\gamma_1-\gamma_2$. Stokes' theorem and the flatness condition in Eq.~(\ref{eom_3d_BF}) give $Q_e(\gamma_1)-Q_e(\gamma_2)=\int_{\partial\Sigma}\frac{N}{2\pi}a=\int_{\Sigma}\frac{N}{2\pi}da=0$. An identical argument applies to $Q_m(\sigma)$ using $\frac{N}{2\pi}db=0$.

Therefore, we can construct two gauge invariant topological operators that generate two higher-form symmetries,
\begin{align}
		U_{q_{e}}\left(\gamma\right)
		= & e^{{\rm i}\frac{2 \pi q_{e}}{N}Q_{e}\left(\gamma\right)}
		= \exp\left({\rm i}q_{e}\int_{\gamma}a\right),\nonumber\\
		U_{q_{m}}\left(\sigma\right)
		= & e^{ {\rm i} \frac{2 \pi q_{m}}{N} Q_{m}\left(\sigma\right)} = \exp\left({\rm i}q_{m}\int_{\sigma}b\right).
	\end{align}
    The compact $BF$ path integral imposes $\oint a\in\frac{2\pi}{N}\mathbb{Z} \mod 2\pi$, which in turn gives $Q_{e}\left(\gamma\right)\in\mathbb{Z} \mod N$, meaning that $q_{e}$ is equivalent to $q_{e}+N$. Furthermore, large-gauge invariance under $\int_{\gamma}a\rightarrow\int_{\gamma}a+2\pi$ requires $q_{e}$ to be an integer. Consequently, $q_{e}$ takes values in $\mathbb{Z}_{N}$. The symmetry operators satisfy
	\begin{equation}
	U_{q_{e}}\left(\gamma\right)\times U_{q_{e}'}\left(\gamma\right)
	= U_{q_{e}+q_{e}'}\left(\gamma\right),
	\qquad
	U_{N}\left(\gamma\right)=1.
	\end{equation}
Similarly, $q_{m} \in \mathbb{Z}_{N}$ and $U_{q_{m}}\left(\sigma\right)$ has a similar algebra. Counting the codimensions of the supporting submanifolds, we know that the operators $U_{q_{e}}\left(\gamma\right)$ and $U_{q_{m}}\left(\sigma\right)$ implement a $\mathbb{Z}_{N}$ $2$-form symmetry and a $\mathbb{Z}_{N}$ $1$-form symmetry, respectively.

We now explain the physical picture of these symmetry operators. From the correspondence between gauge theory and topological order, the operator $e^{{\rm i}\int_{\gamma}a}$ is the Wilson operator associated with a particle excitation carrying one unit of gauge charge, and $\gamma$ can be interpreted as the worldline of this particle. Acting with $e^{{\rm i}\int_{\gamma}a}$ on a state amounts to transporting the particle along $\gamma$ in spacetime. This process probes the gauge flux carried by loop excitations piercing a surface bounded by $\gamma$. We can also think the EoM plays the role of a continuity equation and formally introduce a current analogue $\mathcal{J}_{e} = - * \frac{N}{2\pi}a$ so that $d\left(*\mathcal{J}_{e}\right) = 0$. Since $a=a_{\mu}dx^{\mu}$, this current analogue is a $3$-form, $\mathcal{J}_{e}=-\frac{1}{3!}\frac{N}{2\pi}\epsilon^{\mu \nu \lambda \rho}a_{\mu}\, dx^{\nu}\wedge dx^{\lambda}\wedge dx^{\rho}$. For comparison, in a $(3+1)$D system with a $0$-form $U\left(1\right)$ symmetry, the particle current $j=j_{\mu}dx^{\mu}$ is a $1$-form current. Taking $M_3$ to be the entire $3$d spatial manifold, $\int_{M_{3}} *j$ integrates the particle-number density $j_{0}$ over $M_3$, yielding the total particle number. By direct analogy, $\int_{\gamma} * \mathcal{J}_{e}$ can be interpreted as a generalized ``symmetry charge'' associated with the submanifold $\gamma$. Recall that $\int_{\gamma} * \mathcal{J}_{e}$ measures the rescaled gauge flux encircled by $\gamma$; for a unit-flux loop excitation, it equals the oriented linking number modulo $N$. Equivalently, it can be computed by counting the intersections between $\gamma$ and a three-dimensional volume bounded by $\sigma$ in spacetime. In this sense, $\mathcal{J}_{e}$ formally describes the flow of two-dimensional objects in $(3+1)$D spacetime, in close analogy with $j$ describing the flow of point-like particles. These two-dimensional objects are precisely the worldsheets of loop excitations. The physical interpretation of $Q_m\left(\sigma\right)$ follows in a completely analogous manner.

    When acting on the Hilbert space at a fixed time, these two symmetry operators are supported on $\gamma$ and $\sigma$ within the same $3$d spatial slice $M_3$ and obey the commutation relation \cite{gaiotto_generalized_2015}
	\begin{equation}
		U_{q_{e}}\left(\gamma\right)U_{q_{m}}\left(\sigma\right)
		= e^{\frac{{\rm i}2\pi q_{e}q_{m}}{N}{\rm Int}\left(\gamma,\sigma\right)}
		U_{q_{m}}\left(\sigma\right)U_{q_{e}}\left(\gamma\right),
		\label{eq_BF_equal_time}
	\end{equation}
	where ${\rm Int}\left(\gamma,\sigma\right)$ is the oriented intersection number between $\gamma$ and $\sigma$ in $M_3$. In this sense, $U_{q_{m}}\left(\sigma\right)$ is the charged operator of the symmetry generated by $U_{q_{e}}\left(\gamma\right)$, and vice versa. Take the $3$d $\mathbb{Z}_{2}$ toric code on a $3$-torus with $q_e=q_m=1$ as an example: $U_{q_{e}}\left(\gamma\right)$ corresponds to the closed string ($\gamma$) of Pauli $Z$, and $U_{q_{m}}\left(\sigma\right)$ corresponds to the closed dual surface ($\sigma$) of Pauli $X$. When $\gamma$ and $\sigma$ intersect an odd number of times, these two operators anti-commute. 
    
    As spacetime defects, these operators obey Abelian fusion rules, while their mutual braiding is encoded in a linking-dependent correlation function. For disjoint closed supports in $\mathbb{R}^{4}$, the correlation function normalized by the vacuum partition function is
    \begin{equation}
        \left\langle U_{q_{e}}\left(\gamma\right)U_{q_{m}}\left(\sigma\right) \right\rangle = e^{-\frac{{\rm i}2\pi q_{e}q_{m}}{N}{\rm Lk}\left(\gamma,\sigma\right)},
        \label{eq_BF_spacetime_linking}
    \end{equation}
    where ${\rm Lk}\left(\gamma,\sigma\right)$ is the oriented linking number between $\gamma$ and $\sigma$ in four-dimensional spacetime.

	This example of $\mathbb{Z}_{N}$ $BF$ theory illustrates that a $p$-form symmetry, much like a $0$-form symmetry, originates from a topological operator whose topological invariance is guaranteed by the EoM. Finally, we note that the $BF$ term in Eq.~(\ref{eqn_bftermN}) can be generalized to a multi-component theory with a matrix-valued coefficient~\cite{ye16a,toeplitz2025}. For a finite number of compact gauge fields, an integer coupling matrix can be brought to Smith normal form by independent $GL(n,\mathbb{Z})$ transformations of the $a$ and $b$ fields.

	\subsection{Continuum topological field theory of Borromean-Rings topological order \label{subsec_action_aab}}
	
	The BR topological order~\cite{yp18prl,zhang2021compatible,zhang_non-abelian_2023,huang2025bridge} can be effectively captured by a topological field theory with gauge group
	$G=\prod_{i=1}^{3}\mathbb{Z}_{N_{i}}$, whose action takes the form
	\begin{equation}
		S_{\text{BR}}=\int\sum_{i=1}^{3}\frac{N_{i}}{2\pi}b^{i}  da^{i}+qa^{1}  a^{2}  b^{3},
		\label{eq_aab_action}
	\end{equation}
	where $a^{i}$ and $b^i$ are $1$-form and $2$-form gauge fields, respectively. The first term consists of three conventional $BF$ terms, describing three decoupled $\mathbb{Z}_N$ ($N=N_1,N_2,N_3$) topological orders. The second term is a twisted term involving two $1$-form gauge fields ($a^1,a^2$) and one $2$-form gauge field ($b^3$), which couples the three topological orders. Its coefficient is $q=\frac{pN_1N_2N_3}{(2\pi)^2N_{123}}$, where $N_{123}$ is the greatest common divisor (GCD) of $N_1,N_2,N_3$, and $p\in\mathbb{Z}_{N_{123}}$ labels the twist. Unless stated otherwise, the discussion below assumes $p\ne0$ in $\mathbb{Z}_{N_{123}}$. For $p=0$, the theory reduces to the untwisted $BF$ theory and its fundamental Wilson operators are invertible.
	
	The three $BF$ terms impose flatness conditions on
	$a^{1},a^{2},b^{3}$, leading to the gauge transformations
	$a^{1,2}\rightarrow a^{1,2}+d\chi^{1,2}$ and
	$b^{3}\rightarrow b^{3}+dV^{3}$.
	In contrast, the twisted term $a^{1}a^{2}b^{3}$ induces nontrivial shift
	contributions to the gauge transformations of $b^{1},b^{2}$, and $a^{3}$.
	Explicitly, these gauge transformations take the form
	\begin{align}
		a^{3}\rightarrow & a^{3}+d\chi^{3}
		-\frac{2\pi q}{N_{3}}\left(\chi^{1}a^{2}+\frac{1}{2}\chi^{1}d\chi^{2}\right)\nonumber \\
		& +\frac{2\pi q}{N_{3}}\left(\chi^{2}a^{1}+\frac{1}{2}\chi^{2}d\chi^{1}\right),\\
		b^{1}\rightarrow & b^{1}+dV^{1}
		-\frac{2\pi q}{N_{1}}\left(\chi^{2}b^{3}-a^{2}V^{3}+\chi^{2}dV^{3}\right),\\
		b^{2}\rightarrow & b^{2}+dV^{2}
		+\frac{2\pi q}{N_{2}}\left(\chi^{1}b^{3}-a^{1}V^{3}+\chi^{1}dV^{3}\right).
	\end{align}
	
	To identify the topological operators,
	we first derive the EoMs by requiring the action to remain invariant under infinitesimal variations of the gauge fields.
	For instance, under $b^{1}\rightarrow b^{1}+\delta b^{1}$,
	the condition $\delta S\left[\delta b^{1}\right]=0$
	implies $\frac{N_{1}}{2\pi}da^{1}=0$, which is the EoM of the gauge field $b^{1}$.
	The full set of EoMs together with the corresponding topological operators of $S_{\text{BR}}$ is summarized in
	Table~\ref{tab_EOM_aab}. Among these symmetry operators, some are invertible whereas others are
	non-invertible.
	In particular, due to the presence of the twisted term,
	there exist two distinct classes of EoMs, which we refer to as type-I and type-II, respectively,
	and which will be discussed in detail in the following subsections.
	
	\begin{table*}
		\renewcommand{\arraystretch}{2.2}  
		\caption{Equations of motion, topological/symmetry operators, and corresponding
representations of the operators in the theory $S_{\text{BR}}$. Here,
$e_{i}\in\mathbb{Z}_{N_{i}}$, $r_{i} \in \mathbb{Z}_{N_{i}} $, $K_{1}=\frac{pN_{2}N_{3}}{N_{123}}$, $K_{2}=\frac{pN_{1}N_{3}}{N_{123}}$,
$K_{3}=\frac{pN_{1}N_{2}}{N_{123}}$. The factors $d$ in the last three topological operators are $d_{r_{1}00}=\frac{N_{123}}{\gcd\left(r_{1}p,N_{123}\right)}$,
$d_{0r_{2}0}=\frac{N_{123}}{\gcd\left(r_{2}p,N_{123}\right)}$, $d_{00r_{3}}=\frac{N_{123}}{\gcd\left(r_{3}p,N_{123}\right)}$,
known as the quantum dimensions. These topological operators
are supported on closed submanifolds, $1$-dimensional $\gamma$ and
$2$-dimensional $\sigma$. Here, $\gamma$ is a closed circle and $\sigma$ is a torus
in spacetime. They correspond to the worldline and worldsheet of particle
and loop excitations. We can use $\tau$ and $\left(\eta,\tau\right)$
to parameterize $\gamma$ and $\sigma$, i.e., $\sigma=S_{\eta}^{1}\times S_{\tau}^{1}$.
The notations $\widetilde{\phi}^{1},\widetilde{\phi}^{2},\widetilde{\beta}^{3}$
are defined as $\widetilde{\phi}^{1}\left(\tau\right)=\protect\int_{\tau_{0}}^{\tau}a_{\tau}^{1}\left(\mu\right)d\mu$,
$\widetilde{\phi}^{2}\left(\tau\right)=\protect\int_{\tau_{0}}^{\tau}a_{\tau}^{2}\left(\mu\right)d\mu$,
$\widetilde{\beta}^{3}\left(\tau\right)=\int_{\tau_{0}}^{\tau}d\mu\left(\int_{S_{\eta}^{1}}b_{\tau\eta}^{3}\left(\mu,\eta\right)d\eta\right)$.
They are cumulative integrals of gauge fields on the supporting submanifolds. We construct these topological operators from the
topological invariance guaranteed by the EoMs and the gauge invariance
required by the theory, see Sec.~\ref{subsec_special_type_EoM_line} and \ref{subsec_special_type_EoM_surface}. The first three topological
operators generate invertible $2$-form, $2$-form, and $1$-form
symmetries, respectively. The last three topological operators are
non-invertible when the quantum dimension $d>1$, as they are associated with nontrivial projective representations
(denoted as ``proj. rep.'' in the table), see Sec.~\ref{subsec_special_type_EoM_line}, \ref{subsec_special_type_EoM_surface}, and \ref{subsec_fusion_rule_algebraic}.\label{tab_EOM_aab}}
		
		\begin{tabular*}{\textwidth}{@{\extracolsep{\fill}}cccc}
			\hline 
EoMs & Topological/symmetry operators & Associated representations\tabularnewline
\hline 
$\frac{N_{1}}{2\pi}da^{1}=0$ & $U_{e_{1}}\left(\gamma\right)=\exp\left({\rm i}e_{1}\int_{\gamma}a^{1}\right)$ & $1$D representation of $\mathbb{Z}_{N_{1}}$\tabularnewline
$\frac{N_{2}}{2\pi}da^{2}=0$ & $U_{e_{2}}\left(\gamma\right)=\exp\left({\rm i}e_{2}\int_{\gamma}a^{2}\right)$ & $1$D representation of $\mathbb{Z}_{N_{2}}$\tabularnewline
$\frac{N_{3}}{2\pi}db^{3}=0$ & $U_{e_{3}}\left(\sigma\right)=\exp\left({\rm i}e_{3}\int_{\sigma}b^{3}\right)$ & $1$D representation of $\mathbb{Z}_{N_{3}}$\tabularnewline
$\frac{N_{1}}{2\pi}db^{1}+qa^{2}b^{3}=0$ & $\begin{array}{cc}
L_{r_{1}00}\left(\sigma\right)= & d_{r_{1}00}\times\exp\left[{\rm i}r_{1}\int_{\sigma}b^{1}+\frac{{\rm i}r_{1}K_{1}}{4\pi}\left(\int_{\sigma}b^{3}\widetilde{\phi}^{2}-\int_{S_{\tau}^{1}}a_{\tau}^{2}\widetilde{\beta}^{3}\right)\right]\\
 & \times\delta\left(\frac{r_{1}pN_{3}}{N_{123}}\int_{\sigma}b^{3}\right)\delta\left(\frac{r_{1}pN_{2}}{N_{123}}\int_{S_{\tau}^{1}}a_{\tau}^{2}\right)
\end{array}$ & $d_{r_{1} 00}$D proj. rep. of $\mathbb{Z}_{N_{2}}\times\mathbb{Z}_{N_{3}}$\tabularnewline
$\frac{N_{2}}{2\pi}db^{2}-qa^{1}b^{3}=0$ & $\begin{array}{cc}
L_{0r_{2}0}\left(\sigma\right)= & d_{0r_{2}0}\times\exp\left[{\rm i}r_{2}\int_{\sigma}b^{2}-\frac{{\rm i}r_{2}K_{2}}{4\pi}\left(\int_{\sigma}b^{3}\widetilde{\phi}^{1}-\int_{S_{\tau}^{1}}a_{\tau}^{1}\widetilde{\beta}^{3}\right)\right]\\
 & \times\delta\left(\frac{r_{2}pN_{3}}{N_{123}}\int_{\sigma}b^{3}\right)\delta\left(\frac{r_{2}pN_{1}}{N_{123}}\int_{S_{\tau}^{1}}a_{\tau}^{1}\right)
\end{array}$ & $d_{0 r_{2} 0}$D proj. rep. of $\mathbb{Z}_{N_{1}}\times\mathbb{Z}_{N_{3}}$\tabularnewline
$\frac{N_{3}}{2\pi}da^{3}+qa^{1}a^{2}=0$ & $\begin{array}{cc}
L_{00r_{3}}\left(\gamma\right)= & d_{00r_{3}}\times\exp\left[{\rm i}r_{3}\int_{\gamma}a^{3}+{\rm i}\int_{\gamma}\frac{r_{3}K_{3}}{4\pi}\left(\widetilde{\phi}^{1}a^{2}-\widetilde{\phi}^{2}a^{1}\right)\right]\\
 & \times\delta\left(\frac{r_{3}pN_{1}}{N_{123}}\int_{\gamma}a^{1}\right)\delta\left(\frac{r_{3}pN_{2}}{N_{123}}\int_{\gamma}a^{2}\right)
\end{array}$ & $d_{00r_{3}}$D proj. rep. of $\mathbb{Z}_{N_{1}}\times\mathbb{Z}_{N_{2}}$\tabularnewline
\hline 
	\end{tabular*}
	\end{table*}
	
	\subsection{Type-I EoMs and invertible symmetry operators
		\label{subsec_normal_type_EoM}}
    
	We begin by considering a local variation of the gauge field $b^1$.
	Under $b^1 \rightarrow b^1 + \delta b^1$, the variation of the action is
	\[
	\delta S \left[\delta b^1\right]
	= \int_{M^4} \frac{N_1}{2\pi} \,\delta b^1 \, da^1 .
	\]
	Requiring $\delta S \left[\delta b^1 \right] = 0$ for an arbitrary
	infinitesimal perturbation $\delta b^1$, we obtain the EoM
	\begin{equation}
		\label{eq_BR_EoM_a1}
		\frac{N_1}{2\pi} \, da^1 = 0.
	\end{equation}
This EoM states that the gauge field is flat. We refer to equations of this form as type-I EoMs; their vanishing exterior derivatives ensure the topological invariance of the corresponding Wilson operators.

	The EoM~(\ref{eq_BR_EoM_a1}) immediately gives rise to a gauge invariant topological operator
    \begin{equation}
		U_{e_1} \left(\gamma\right)
		= e^{{\rm i} \frac{2\pi e_1}{N_{1}}\cdot \frac{N_{1}}{2\pi}\int_{\gamma}a^1}=\exp\left({\rm i} e_1 \int_{\gamma} a^1\right).
		\label{eq_symm_op_a1}
	\end{equation}
    The topological invariance comes from the fact that the quantity $Q_{1,\gamma} = \frac{N_{1}}{2\pi}\int_{\gamma} a^1$ depends only on the homology class of the closed curve $\gamma$: when $\gamma$ becomes $\gamma'$ under a smooth deformation, $Q_{1,\gamma}-Q_{1,\gamma '}
	= \int_{\gamma - \gamma '} \frac{N_{1}}{2\pi}a^{1}
	= \int_{\Sigma} \frac{N_{1}}{2\pi}d a^{1}=0$,
	where $\Sigma$ is a surface with $\partial\Sigma=\gamma-\gamma'$ and we
	have used the Stokes' theorem. 
    
    The object charged under this symmetry is the closed worldsheet of the
	loop excitation described by $b^1$.
	Correspondingly, $Q_{1,\gamma}$ counts the linking number of $\gamma$ and the loop worldsheet. The compact gauge-field path integral also imposes the holonomy quantization
	$\oint_{\gamma} a^1 = \frac{2\pi k_1}{N_1}$ with
	$k_1 \in \mathbb{Z}_{N_{1}}$.
	Equivalently, the flux of $a^1$ piercing a surface bounded by $\gamma$ is
	quantized in integer multiples of $\frac{2\pi}{N_1}$.
	Imposing large-gauge invariance together with the quantization of
	$\int_{\gamma} a^1$, we find that $e_1$ is an integer defined modulo
	$N_1$.
	As a result, the topological operator
	$U_{e_1} \left(\gamma\right)$ with
	$e_1 \in \left\{0,1,\cdots,N_1-1\right\}$ implements a
	$\mathbb{Z}_{N_1}$ $2$-form symmetry.

    Following the same reasoning, by varying the gauge
	fields $b^{2}$ and $a^{3}$,
	\begin{equation}
		\label{eq_BR_EoM_a2_b3}
		\frac{N_{2}}{2\pi}da^{2}=0,\quad\frac{N_{3}}{2\pi}db^{3}=0,
	\end{equation}
    we identify corresponding topological operators (symmetry operators)
    \begin{equation}
		U_{e_{2}}\left(\gamma\right)
		= e^{{\rm i}e_{2}\int_{\gamma}a^{2}},\ U_{e_{3}}\left(\sigma\right)
		= e^{{\rm i}e_{3}\int_{\sigma}b^{3}},
	\end{equation}
    where
	$e_{2}\in\left\{0,1,\cdots,N_{2}-1\right\}$
	and
	$e_{3}\in\left\{0,1,\cdots,N_{3}-1\right\}$.
	We note that the symmetry operators
	$U_{e_{1}}\left(\gamma\right)$,
	$U_{e_{2}}\left(\gamma\right)$, and
	$U_{e_{3}}\left(\sigma\right)$
	coincide with the Wilson operators of Abelian particle and loop
	excitations.
	The operator $U_{e_{2}}\left(\gamma\right)$ implements a
	$\mathbb{Z}_{N_{2}}$ $2$-form symmetry, while
	$U_{e_{3}}\left(\sigma\right)$ implements a
	$\mathbb{Z}_{N_{3}}$ $1$-form symmetry, in the sense that
	$U_{e_{i}}\times U_{e_{i}'}=U_{e_{i}+e_{i}'}$ and
	$U_{N_{i}} = \mathbbm{1}$.
	Thus far, we have identified a basic family of invertible symmetry operators in the effective field theory description of the BR topological order. Additional invertible operators can arise from suitable charge multiples in the type-II sectors, as discussed in Sec.~\ref{subsec_general_symmetry_operators}.

	\subsection{Type-II EoM and non-invertible topological line operators\label{subsec_special_type_EoM}\label{subsec_special_type_EoM_line}} 
    
	Besides the type-I EoMs discussed above, the theory also
	admits EoMs in which the exterior derivative of a gauge field is sourced
	by the twist. The corresponding Wilson factors require an auxiliary
	system to define gauge invariant topological operators. We call them type-II EoMs. The corresponding topological operators are equipped with additional structures: in $S_{\text{BR}}$, they lead to a class of non-invertible topological line operators (this section) and two classes of non-invertible surface operators (Sec.~\ref{subsec_special_type_EoM_surface}). 
    
    Varying $b^{3}$ gives
\begin{equation}
\frac{N_{3}}{2\pi}da^{3}+qa^{1}a^{2}=0.\label{eq_BR_EoM_a3}
\end{equation}
In contrast to the EoMs (\ref{eq_BR_EoM_a1}) and (\ref{eq_BR_EoM_a2_b3}), the above EoM
is not simply a vanishing exterior derivative. One can indeed
construct an operator like $e^{{\rm i}\left(\int_{\gamma}a^{3}+\int_{\Sigma}\frac{2\pi q}{N_{3}}a^{1}a^{2}\right)}$
with $\partial\Sigma=\gamma$, which is invariant under a smooth deformation
of the surface $\Sigma$, but it is not a genuine
topological line operator since it depends on the open surface bounded by $\gamma$. On the other hand,
the conventional Wilson-type operator $e^{{\rm i}\int_{\gamma}a^{3}}$
is indeed supported only on $\gamma$ but it is not gauge invariant,
making it impossible to be a legitimate topological operator. To resolve
this dilemma, we can stack $e^{{\rm i}\int_{\gamma}a^{3}}$ to a $1$-dimensional auxiliary
system living on $\gamma$ coupled to gauge fields $a^{1}$ and
$a^{2}$. In the language of anomaly matching, this auxiliary system is not gauge invariant
on its own, and its anomaly is canceled by $\int_{\gamma}a^{3}$. This implements the same idea as the defect-dressing mechanism of Refs.~\cite{kaidi_kramers-wannier-like_2022,Shao_non-inv_SM-2022}, in which an auxiliary theory on the defect (operator) cancels its gauge variation. Below we will show the auxiliary system can have a nontrivial
Hilbert space which makes the genuine topological line operators non-invertible. 

The auxiliary $1$-dimensional system coupling to gauge fields
$a^{1},a^{2}$ is described by the action 
\begin{equation}
S_{\text{1d aux}}\left(a^{1},a^{2}\right)=\int_{\gamma}\frac{K_{3}}{2\pi}\left(-\phi^{1}d\phi^{2}+\phi^{1}a^{2}-\phi^{2}a^{1}\right),
\label{eq_S_1d_aux}
\end{equation}
where $K_{3}=\frac{(2\pi)^{2}q}{N_{3}}=\frac{pN_{1}N_{2}}{N_{123}}$, and $\phi^{i}$'s are \textit{compact} scalar
fields, $\phi^{i} \sim \phi^{i} + 2\pi$, living on $\gamma$. Thus, once canonical quantization is performed, quantum operators in the form of $e^{i\phi^1}$ and $e^{i\phi^2}$ become useful. One can check that $e^{{\rm i}\int_{\gamma}a^{3}+{\rm i}S_{\text{1d aux}}\left(a^{1},a^{2}\right)}$
is gauge invariant under the gauge transformations of $a^{i}$ and $\phi^{i}\rightarrow\phi^{i}+\chi^{i}$. The topological line operator is 
\begin{equation}
L_{001}\left(\gamma\right)=e^{{\rm i}\int_{\gamma}a^{3}}\mathcal{Z}_{\text{1d aux}}\left(a^{1},a^{2}\right)\label{eqexplain}
\end{equation}
with $\mathcal{Z}_{\text{1d aux}}\left(a^{1},a^{2}\right)=\int_{\mathcal H}\mathcal{D}\phi^{1}\mathcal{D}\phi^{2}e^{{\rm i}S_{\text{1d aux}}\left(a^{1},a^{2}\right)}$
being the partition function in the irreducible auxiliary sector $\mathcal H$ specified below. The fields $\phi^{1},\phi^{2}$ are integrated only on $\gamma$, and $a^{1},a^{2}$ in this action denote the pullbacks of the bulk fields. Inserting the line into the bulk path integral gives
\begin{equation*}
\begin{aligned}
\left\langle L_{001}(\gamma)\right\rangle
={}&\frac{1}{\mathcal{Z}_{\mathrm{BR}}}
\int\mathcal{D}[a^{i},b^{i}]\,e^{{\rm i}S_{\mathrm{BR}}}\\
&\times e^{{\rm i}\int_{\gamma}a^{3}}\,
\mathcal{Z}_{\text{1d aux}}(a^{1}|_{\gamma},a^{2}|_{\gamma}),
\end{aligned}
\end{equation*}
where $\mathcal{Z}_{\mathrm{BR}}=\int\mathcal{D}[a^{i},b^{i}]e^{{\rm i}S_{\mathrm{BR}}}$. Here and below, the notation $\int_{\mathcal H}$ includes the holonomy rephasing that implements the ordered convention specified in Appendix~\ref{Appendix_line_construction}. Thus $a^{i},b^{i}$ remain the dynamical bulk fields, while the auxiliary fields contribute only on the defect.

Treating $\gamma$ as a ``time'' circle, $\mathcal{Z}_{\text{1d aux}}$ describes a $\left(0+1\right)$D quantum field theory in which $a^1, a^2$ are background fields. The key question is
how their holonomies $\int_{\gamma}a^{i}=\frac{2\pi k_{i}}{N_{i}}$,
$k_{i}\in\mathbb{Z}_{N_{i}}$, act on the auxiliary fields. After
canonical quantization (see Appendix~\ref{Appendix_line_construction}),
the holonomy of $a^{1}$ ($a^{2}$) acts as $U_{1}^{k_{1}}$ ($U_{2}^{k_{2}}$),
with $U_{1}=e^{-\frac{{\rm i}K_{3}}{N_{1}}\phi^{2}}$ and $U_{2}=e^{\frac{{\rm i}K_{3}}{N_{2}}\phi^{1}}$
satisfying 
\begin{equation}
U_{1}^{N_{1}}=U_{2}^{N_{2}}=\mathds{1},\qquad U_{1}U_{2}=e^{-\frac{{\rm i}2\pi p}{N_{123}}}U_{2}U_{1}.\label{eq_line_projective_algebra}
\end{equation}
Thus, the holonomies act projectively on the auxiliary Hilbert space $\mathcal H$
under the gauge subgroup $\mathbb{Z}_{N_{1}}\times\mathbb{Z}_{N_{2}}$. We take $\mathcal{H}$ to be the minimal irreducible projective representation space 
of Eq.~(\ref{eq_line_projective_algebra}),\footnote{The term $-\frac{K_{3}}{2\pi}\int_{\gamma}\phi^{1}d\phi^{2}$
alone admits a $K_{3}$-dimensional Hilbert space, whose representation under the holonomies can be reducible. The elementary topological operator considered here is defined by taking a single minimal irreducible sector $\mathcal H$; this choice is indicated by the subscript on the auxiliary functional integral. A trace over the full auxiliary Hilbert space would instead give the corresponding direct sum of topological operators.} whose dimension $d=\frac{N_{123}}{\gcd\left(p,N_{123}\right)}$
is the order of the phase $e^{-\frac{{\rm i}2\pi p}{N_{123}}}$. A
convenient basis is the clock-shift basis $\left\{ |n\rangle\right\} _{n=0}^{d-1}$
with
\begin{equation}
U_{1}|n\rangle=e^{-\frac{{\rm i}2\pi pn}{N_{123}}}|n\rangle,\qquad U_{2}|n\rangle=|n+1\rangle.\label{eq_line_clock_basis}
\end{equation}
Evaluating $\mathcal{Z}_{\text{1d aux}}$ as a trace over $\mathcal{H}$
(see Appendix~\ref{Appendix_line_construction}), we obtain 
\begin{align}
L_{001}\left(\gamma\right)= & \,d\cdot e^{{\rm i}\int_{\gamma}a^{3}+{\rm i}\Phi}\nonumber \\
 & \times\delta\left(\frac{pN_{1}}{N_{123}}\int_{\gamma}a^{1}\right)\delta\left(\frac{pN_{2}}{N_{123}}\int_{\gamma}a^{2}\right),\label{eq_L001}
\end{align}
where $\delta\left(x\right)=1$ if $x\in2\pi\mathbb{Z}$ and $\delta\left(x\right)=0$
otherwise, and $\Phi=\frac{K_{3}}{4\pi}\int_{\gamma}\left(\widetilde{\phi}^{1}a^{2}-\widetilde{\phi}^{2}a^{1}\right)$
with $\widetilde{\phi}^{i}\left(\tau\right)=\int_{\tau_{0}}^{\tau}a^{i}\left(\mu\right)d\mu$
is a $c$-number phase. The delta functions restrict the holonomies
to $k_{1},k_{2}=0\mod d$.

The mathematical meaning of the topological line operator supported
on $\gamma$ is clear: it has a $d$-dimensional projective representation
space of $\mathbb{Z}_{N_{1}}\times\mathbb{Z}_{N_{2}}$. Therefore,
the fusion of two such operators corresponds to the product of two representation
spaces, which decomposes into a sum of other representation spaces,
i.e., other topological line operators. In this sense, the topological
line operator $L_{001}\left(\gamma\right)$ is non-invertible when
$d>1$ and it generates a non-invertible symmetry. Physically, it corresponds to a non-Abelian particle excitation
in the BR topological order, whose quantum dimension is $d$.

\subsection{Type-II EoMs and non-invertible topological surface operators\label{subsec_special_type_EoM_surface}}
Varying $a^{1}$ and $a^{2}$ gives the type-II EoMs for $b^{1}$ and $b^{2}$. Consider first
\begin{equation}
\frac{N_{1}}{2\pi}db^{1}+qa^{2}b^{3}=0.\label{eq_BR_EoM_b1}
\end{equation}
As in Sec.~\ref{subsec_special_type_EoM_line}, $e^{{\rm i}\left(\int_{\sigma}b^{1}+\int_{\Omega}\frac{2\pi q}{N_{1}}a^{2}b^{3}\right)}$
with $\partial\Omega=\sigma$ is gauge invariant but depends on the
open $3$-manifold $\Omega$, while the conventional Wilson-type operator $e^{{\rm i}\int_{\sigma}b^{1}}$
on a closed surface $\sigma$ is not gauge invariant. We therefore
stack $e^{{\rm i}\int_{\sigma}b^{1}}$ with a $2$-dimensional auxiliary
system on $\sigma$,
\begin{equation}
S_{\text{2d aux}}\left(a^{2},b^{3}\right)=\int_{\sigma}\frac{K_{1}}{2\pi}\left(-\phi^{2}d\beta^{3}+\beta^{3}a^{2}+\phi^{2}b^{3}\right),\label{eq_S_2d_aux}
\end{equation}
where $\sigma$ is a torus in spacetime, $K_{1}=\frac{(2\pi)^{2}q}{N_{1}}=\frac{pN_{2}N_{3}}{N_{123}}$,
the $0$-form $\phi^{2}$ and $1$-form $\beta^{3}$ are compact auxiliary
fields living on $\sigma$.  We can check that $e^{{\rm i}\int_{\sigma}b^{1}+{\rm i}S_{\text{2d aux}}\left(a^{2},b^{3}\right)}$
is invariant under the gauge transformations of $b^1, a^{2}, b^{3}$ and $\beta^{3}\rightarrow\beta^{3}+V^{3}$, $\phi^{2}\rightarrow\phi^{2}+\chi^{2}$. The topological surface operator is
\begin{equation}
L_{100}\left(\sigma\right)=e^{{\rm i}\int_{\sigma}b^{1}}\mathcal{Z}_{\text{2d aux}}\left(a^{2},b^{3}\right)
\end{equation}
with $\mathcal{Z}_{\text{2d aux}}\left(a^{2},b^{3}\right)=\int_{\mathcal H}\mathcal{D}\phi^{2}\mathcal{D}\beta^{3}e^{{\rm i}S_{\text{2d aux}}\left(a^{2},b^{3}\right)}$
being the partition function in the irreducible auxiliary sector $\mathcal H$ specified below. Here the auxiliary fields are integrated only on $\sigma$, and $a^{2},b^{3}$ denote the pullbacks of the bulk fields. As for the line insertion, this auxiliary partition function is inserted into the path integral over $a^{i},b^{i}$ with bulk action $S_{\mathrm{BR}}$.

We parameterize the surface as $\sigma = S^{1}_{\eta} \times S^{1}_{\tau}$, with $\eta$ ($\tau$) as the ``spatial'' (``time'') direction on $\sigma$. For configurations with $a_{\eta}^{2}=0$
and $a_{\tau}^{2}$ independent of $\eta$, $\phi^{2}$ depends only
on $\tau$, and $S_{\text{2d aux}}$ reduces to a $\left(0+1\right)$D
system of $\phi^{2}$ and the spatial holonomy $B^{3}=\int_{S_{\eta}^{1}}\beta^{3}$
(see Appendix~\ref{Appendix_surface_construction}). The holonomies
$\int_{S_{\tau}^{1}}a^{2}=\frac{2\pi k_{2}}{N_{2}}$ and $\int_{\sigma}b^{3}=\frac{2\pi k_{3}}{N_{3}}$
then act as $U_{2}^{k_{2}}$ and $U_{3}^{k_{3}}$, with $U_{2}=e^{-\frac{{\rm i}K_{1}}{N_{2}}B^{3}}$
and $U_{3}=e^{\frac{{\rm i}K_{1}}{N_{3}}\phi^{2}}$ satisfying
\begin{equation}
U_{2}^{N_{2}}=U_{3}^{N_{3}}=\mathds{1},\qquad U_{2}U_{3}=e^{-\frac{{\rm i}2\pi p}{N_{123}}}U_{3}U_{2}.\label{eq_surface_projective_algebra}
\end{equation}
This algebra is in the same spirit as Eq.~(\ref{eq_line_projective_algebra}):
the holonomies of $a^{2}$ and $b^{3}$ act projectively on the auxiliary
Hilbert space under $\mathbb{Z}_{N_{2}}\times\mathbb{Z}_{N_{3}}$. The
Hilbert space is again the $d$-dimensional projective representation space,
$d=\frac{N_{123}}{\gcd\left(p,N_{123}\right)}$, with clock-shift
basis $U_{2}|n\rangle=e^{-\frac{{\rm i}2\pi pn}{N_{123}}}|n\rangle$
and $U_{3}|n\rangle=|n+1\rangle$. Evaluating $\mathcal{Z}_{\text{2d aux}}$
as a trace (see Appendix~\ref{Appendix_surface_construction}), we
obtain
\begin{align}
L_{100}\left(\sigma\right)= & d\times e^{{\rm i}\int_{\sigma}b^{1}+\frac{{\rm i}K_{1}}{4\pi}\left(\int_{\sigma}b^{3}\widetilde{\phi}^{2}-\int_{S_{\tau}^{1}}a_{\tau}^{2}\widetilde{\beta}^{3}\right)}\nonumber \\
 & \times\delta\left(\frac{pN_{3}}{N_{123}}\int_{\sigma}b^{3}\right)\delta\left(\frac{pN_{2}}{N_{123}}\int_{S_{\tau}^{1}}a_{\tau}^{2}\right)
 \label{eq_L100}
\end{align}
where $\widetilde{\phi}^{2}\left(\tau\right)=\int_{\tau_{0}}^{\tau}a_{\tau}^{2}\left(\mu\right)d\mu$
and $\widetilde{\beta}^{3}\left(\tau\right)=\int_{\tau_{0}}^{\tau}d\mu\int_{S_{\eta}^{1}}b_{\tau\eta}^{3}\left(\mu,\eta\right)d\eta$,
with $S_{\tau}^{1}$ parameterized by $\tau\in\left[\tau_{0},\tau_{f}\right)$.

Similarly, the type-II EoM of gauge field $b^{2}$,
\begin{equation}
\frac{N_{2}}{2\pi}db^{2}-qa^{1}b^{3}=0,
\end{equation}
leads to (see Appendix~\ref{Appendix_surface_construction})
\begin{align}
L_{010}\left(\sigma\right)= & d\times e^{{\rm i}\int_{\sigma}b^{2}-\frac{{\rm i}K_{2}}{4\pi}\left(\int_{\sigma}b^{3}\widetilde{\phi}^{1}-\int_{S_{\tau}^{1}}a_{\tau}^{1}\widetilde{\beta}^{3}\right)}\nonumber \\
 & \times\delta\left(\frac{pN_{3}}{N_{123}}\int_{\sigma}b^{3}\right)\delta\left(\frac{pN_{1}}{N_{123}}\int_{S_{\tau}^{1}}a_{\tau}^{1}\right)
 \label{eq_L010}
\end{align}
where $d=\frac{N_{123}}{\gcd\left(N_{123},p\right)}$, $K_{2}=\frac{pN_{1}N_{3}}{N_{123}}$, and $\widetilde{\phi}^{1}\left(\tau\right)=\int_{\tau_{0}}^{\tau}a_{\tau}^{1}\left(\mu\right)d\mu$.

Similar to the topological line operator discussed in the previous
section, the topological surface operator $L_{100}\left(\sigma\right)$
[$L_{010}\left(\sigma\right)$] is associated with a $d$-dimensional projective
representation of $\mathbb{Z}_{N_{2}}\times\mathbb{Z}_{N_{3}}$ [$\mathbb{Z}_{N_{1}}\times\mathbb{Z}_{N_{3}}$], and is non-invertible when $d>1$. Physically,
such non-invertible topological surface operators correspond to the non-Abelian loop
excitations in the $\left(3+1\right)$D topological order described
by $S_{\text{BR}}$. The integer factor $d$ of the topological surface
operator is the quantum dimension of the corresponding non-Abelian
loop excitation. 

So far, in this section and Sec.~\ref{subsec_special_type_EoM_line}, we have constructed the fundamental topological operators in the theory $S_{\text{BR}}$, see Table~\ref{tab_EOM_aab}. The guiding principles are topological invariance guaranteed by the EoMs and gauge invariance required by the theory itself. Such topological operators implement invertible and non-invertible higher-form symmetries. These topological operators have a clear physical picture that they correspond to the topological excitations in the BR topological order. Inserting a topological operator means creating and annihilating an excitation whose trajectory in spacetime is the submanifold supporting the topological operator.

\section{Fusion rules of symmetry operators
\label{sec_fusion_rule}}

After revealing the basic topological operators in $S_{\text{BR}}$, the next step is to study the fusion rules of them, which are important data characterizing
the generalized symmetries. First, we identify the complete set of
topological operators. Once the algebraic structure of each individual
symmetry operator is obtained, a fusion is understood from two complementary
viewpoints: (1) it corresponds to taking the tensor product of the
projective Hilbert spaces associated with each operator and decomposing
it into irreducible components; (2) on the other hand, field-theoretically, this
reduces to evaluating the product of the delta functions
which, as projectors, are responsible for the non-invertibility. We will illustrate these ideas with examples. The plan of this section is the following. We first give the general formalism of topological operators which covers all possible topological operators in the theory. Next, we show how to calculate a fusion rule with an example from the algebraic and field-theoretical perspective. Finally, we explain the field-theoretical calculation for the general fusion rules.

\subsection{General formalism of symmetry operators\label{subsec_general_symmetry_operators}}

In Sec.~\ref{subsec_special_type_EoM_line} and \ref{subsec_special_type_EoM_surface}, we have revealed the basic symmetry operators (topological
operators) from the EoMs. In fact, the theory $S_{\text{BR}}$ admits
other symmetry operators. In the following, we will exhaust all of the symmetry operators, which is necessary for the discussion
of the fusion of symmetry operators.

The type-I EoMs lead to the simplest symmetry operators, $U_{e_1}(\gamma)=e^{{\rm i}e_1\int_\gamma a^1}$, $U_{e_2}(\gamma)=e^{{\rm i}e_2\int_\gamma a^2}$, and $U_{e_3}(\sigma)=e^{{\rm i}e_3\int_\sigma b^3}$, with $e_i\in\mathbb{Z}_{N_i}$, $i=1,2,3$. Their product $U_{e_1e_2}(\gamma)=e^{{\rm i}\int_\gamma(e_1a^1+e_2a^2)}$ is also a symmetry operator and corresponds to a particle excitation. The surface $\sigma$ supporting $U_{e_3}(\sigma)$ can be understood as the worldsheet of a loop excitation. We write $\sigma $ as $\sigma=S_{\eta}^1\times S_{\tau}^1$ and let $\gamma$ be the non-contractible circle along the $\tau$ direction on $\sigma$. A composite of these type-I invertible symmetry operators is
\begin{equation}
U_{e_{1}e_{2}e_{3}}\left(\sigma,\gamma\right)=e^{{\rm i}e_{3}\int_{\sigma}b^{3}+{\rm i}\int_{\gamma}\left(e_{1}a^{1}+e_{2}a^{2}\right)}.
\end{equation}
It corresponds to a loop excitation decorated by a particle excitation.
Specifically, $\sigma$ is the worldsheet of the loop excitation,
and $\gamma$ is the worldline of the particle excitation bound to
the loop excitation. 

A more general family of symmetry operators, including both invertible and non-invertible cases, is
\begin{align}
L_{r_{1}r_{2}r_{3}}\left(\sigma,\gamma\right)= & d_{r_{1}r_{2}r_{3}}\times e^{{\rm i}\left(\int_{\sigma}r_{1}b^{1}+\int_{\sigma}r_{2}b^{2}+\int_{\gamma}r_{3}a^{3}\right)}\nonumber\\
& \times e^{{\rm i}\Phi\left(r_{1}, r_{2},r_{3}\right)} 
\times\mathbb{P}\left(r_{1},r_{2},r_{3}\right),
\label{eq_general_L}
\end{align}
where $\sigma$ is a torus denoted as $\sigma = S^{1}_{\eta} \times S^{1}_{\tau}$ and $\gamma$ is the non-contractible path
along the $\tau$-direction on $\sigma$. Physically, $\sigma$ is the
worldsheet of the loop excitation, and $\gamma$ is the worldline
of the particle decorated on that loop. Let us explain the terms in
this expression. First, $r_i\in\mathbb{Z}_{N_i}$, represented by $r_i=0,1,\ldots,N_i-1$.
The integer $d_{r_{1} r_{2} r_{3}}=\frac{N_{123}}{\gcd\left(N_{123},r_{1}p,r_{2}p,r_{3}p\right)}$
is the dimension of the Hilbert space (projective representation space) associated to this operator, also known as the quantum dimension of the operator. We use the ordinary auxiliary trace, with $\mathrm{Tr}_{\mathcal H}\mathbf{1}=d_{r_1r_2r_3}$, which fixes the overall factor in Eq.~(\ref{eq_general_L}) and hence the normalization used in fusion products. The phase factor $e^{{\rm i}\Phi\left(r_{1}, r_{2},r_{3}\right)}$
is: 
\begin{align}
e^{{\rm i}\Phi\left(r_{1},r_{2},r_{3}\right)}= & e^{\frac{{\rm i}r_{1}K_{1}}{4\pi}\left(\int_{\sigma}b^{3}\widetilde{\phi}^{2}-\int_{S^{1}_{\tau}}a_{\tau}^{2}\widetilde{\beta}^{3}\right)}\nonumber \\
 & \times e^{\frac{{\rm i}r_{2}K_{2}}{4\pi}\left(\int_{S^{1}_{\tau}}a_{\tau}^{1}\widetilde{\beta}^{3}-\int_{\sigma}b^{3}\widetilde{\phi}^{1}\right)}\nonumber \\
 & \times e^{\frac{{\rm i}r_{3}K_{3}}{4\pi}\left(\int_{S^{1}_{\tau}}a_{\tau}^{2}\widetilde{\phi}^{1}-\int_{S^{1}_{\tau}}a_{\tau}^{1}\widetilde{\phi}^{2}\right)}
 \label{eq_general_Phi}
\end{align}
where $\widetilde{\phi}^{1}\left(\tau\right)=\int_{\tau_{0}}^{\tau}a_{\tau}^{1}\left(\mu\right)d\mu$,
$\widetilde{\phi}^{2}\left(\tau\right)=\int_{\tau_{0}}^{\tau}a_{\tau}^{2}\left(\mu\right)d\mu$,
$\widetilde{\beta}^{3}\left(\tau\right)=\int_{\tau_{0}}^{\tau}d\mu\left(\int_{S_{\eta}^{1}}b_{\tau\eta}^{3}\left(\mu,\eta\right)d\eta\right)$.
The last term is the projector $\mathbb{P}\left(r_{1},r_{2},r_{3}\right)$
which consists of delta functions:
\begin{align}
\mathbb{P}\left(r_{1},r_{2},r_{3}\right)= & \delta\left(\frac{r_{1}pN_{2}}{N_{123}}\int_{\gamma}a^{2}-\frac{r_{2}pN_{1}}{N_{123}}\int_{\gamma}a^{1}\right)\nonumber \\
 & \times\delta\left(\frac{r_{1}pN_{3}}{N_{123}}\int_{\sigma}b^{3}-\frac{r_{3}pN_{1}}{N_{123}}\int_{\gamma}a^{1}\right)\nonumber \\
 & \times\delta\left(\frac{r_{2}pN_{3}}{N_{123}}\int_{\sigma}b^{3}-\frac{r_{3}pN_{2}}{N_{123}}\int_{\gamma}a^{2}\right).
 \label{eq_general_P}
\end{align}
Within this construction, $L_{r_1r_2r_3}$ is invertible precisely when its quantum dimension equals one:
\begin{align}
d_{r_1r_2r_3}=1
&\quad\Longleftrightarrow\quad N_{123}\mid p r_i\quad(i=1,2,3)\nonumber\\
&\quad\Longleftrightarrow\quad r_i\equiv0\pmod d\quad(i=1,2,3),
\label{eq_invertible_type_II}
\end{align}
where $d=N_{123}/\gcd(p,N_{123})$. Indeed, the allowed holonomy angles then make every argument of $\mathbb{P}$ an integer multiple of $2\pi$, so that $\mathbb{P}=1$, and the auxiliary representation is one-dimensional. The resulting phase operator has an inverse given by its Hermitian conjugate. Thus, products of such $L$ operators with the type-I operators $U_{e_1e_2e_3}$ provide additional invertible operators; the notation $L$ alone does not imply non-invertibility. For $p=0$, the criterion gives $d_{r_1r_2r_3}=1$ for every charge sector, consistent with the untwisted theory.

One can obtain the expression of a general $L_{r_{1}r_{2} r_{3}}\left(\sigma, \gamma\right)$ from the same idea as that in Sec.~\ref{sec_symmetry_operator_aab}. The detailed derivation is collected in Appendix~\ref{Appendix_general_construction}.

\subsection{Example of fusion rule: algebraic viewpoint\label{subsec_fusion_rule_algebraic}}
When constructing topological operators, we have
mentioned a key property of a topological operator: it carries
a representation space of a group. For example, the invertible operator
$U_{e_1}\left(\gamma\right)=e^{{\rm i}\int_{\gamma}e_{1}a^{1}}$
has a $1$-dimensional representation space of $\mathbb{Z}_{N_{1}}$.
On the other hand, the non-invertible topological operator $L_{001}\left(\gamma\right)$
has a $d$-dimensional projective representation space of $\mathbb{Z}_{N_{1}}\times\mathbb{Z}_{N_{2}}$,
with $d=\frac{N_{123}}{\gcd\left(p,N_{123}\right)}$. A fusion of
two topological operators gives the tensor product of the two representation
spaces, and the fusion rule tells its decomposition as a sum of representation spaces
associated with other topological operators. 

The fusion rules of invertible topological operators are group-like:
\begin{align}
 & U_{e_{1}e_{2}e_{3}}\left(\sigma,\gamma\right)\times U_{e_{1}^{\prime}e_{2}^{\prime}e_{3}^{\prime}}\left(\sigma,\gamma\right)\nonumber \\
= & U_{\left(e_{1}+e_{1}^{\prime}\right)\left(e_{2}+e_{2}^{\prime}\right)\left(e_{3}+e_{3}^{\prime}\right)}\left(\sigma,\gamma\right)
\end{align}
where $e_{i}+e_{i}^{\prime}$ is defined modulo $N_{i}$. Since
each invertible operator carries a $1$D representation space, the
fusion of them still produces a $1$D representation space which corresponds
to another invertible operator. 

The fusion of non-invertible topological operators involves the decomposition
of projective representation space. Below we will use an example to
discuss how to calculate the fusion rules. Let us set the gauge group
of our twisted gauge theory $S_{\text{BR}}$ as $G=\mathbb{Z}_{2}\times\mathbb{Z}_{2}\times\mathbb{Z}_{4}$ with $p=1$.
The topological line operator is $L_{00 r_{3}}\left(\gamma\right)$
where $r_{3}=0,1,2,3$. When $r_{3}=1$, the topological line is 
\begin{equation}
L_{001}\left(\gamma\right)=d_{001}\cdot e^{{\rm i}\int_{\gamma}a^{3}+{\rm i}\Phi\left(0,0,1\right)}\cdot\delta\left(\int_{\gamma}a^{1}\right)\delta\left(\int_{\gamma}a^{2}\right)
\end{equation}
with $d_{001}=2$, which carries a projective representation of
gauge group $\mathbb{Z}_{N_{1}}\times\mathbb{Z}_{N_{2}}=\mathbb{Z}_{2}\times\mathbb{Z}_{2}$.
Let $g_{1},g_{2}$ be the generators of $\mathbb{Z}_{2}\times\mathbb{Z}_{2}$
and $\rho_{r_{3}=1}\left(g_{i}\right)$ be the corresponding representation
(i.e., the operator that implements the gauge transformation). Following
the discussion in Sec.~\ref{subsec_special_type_EoM}, we have $\rho_{r_{3}=1}\left(g_{1}\right)\rho_{r_{3}=1}\left(g_{2}\right)=-\rho_{r_{3}=1}\left(g_{2}\right)\rho_{r_{3}=1}\left(g_{1}\right)$,
and the representation space is $2$-dimensional. Similarly, when
$r_{3}=2$, the topological line is 
\begin{equation}
L_{002}\left(\gamma\right)=d_{002}\cdot e^{{\rm i}\int_{\gamma}2\cdot a^{3}+{\rm i}\Phi\left(0,0,2\right)}\cdot\delta\left(2\int_{\gamma}a^{1}\right)\delta\left(2\int_{\gamma}a^{2}\right)
\end{equation}
with $d_{002}=1$, whose representation space for $\mathbb{Z}_2\times\mathbb{Z}_2$ is one-dimensional and satisfies $\rho_{r_3=2}(g_1)\rho_{r_3=2}(g_2)=\rho_{r_3=2}(g_2)\rho_{r_3=2}(g_1)$. Since $\int_\gamma a^i\in\pi\mathbb{Z}$ for $i=1,2$, both projectors in $L_{002}$ are identically one. Thus $L_{002}$ is an invertible line, illustrating an invertible sector beyond the type-I family.

Let us consider the fusion of two $L_{001}\left(\gamma\right)$'s. Recall the picture that a topological operator corresponds to a topological excitation. An $L_{001}\left(\gamma\right)$ corresponds to a non-Abelian particle excitation with $r_{3}=1$ unit of charge. So fusing two such particles should result in a combination of Abelian particles with $r_{3}=2$ units of charge, which in turn correspond to the operator $L_{002}\left(\gamma\right)$. Furthermore, this Abelian particle can have some Abelian particles attached to it, corresponding to the operator $L_{002}\left(\gamma\right)e^{{\rm i}\left(q_{1} \int_{\gamma} a^1 + q_{2} \int_{\gamma} a^2\right)}$ with $q_{i} \in \mathbb{Z}/N_{i}\mathbb{Z}$. When $q_{1}, q_{2}$ take specific values, $L_{002}\left(\gamma\right)e^{{\rm i}\left(q_{1} \int_{\gamma} a^1 + q_{2} \int_{\gamma} a^2\right)}$ and $L_{002}\left(\gamma\right)$ could have the same behavior in a correlation function. However, the delta functions in $L_{002}\left(\gamma\right)$ are identically one. Different $\left(q_{1},q_{2}\right)$'s label different $1$D representations of $\mathbb{Z}_{2}\times\mathbb{Z}_{2}$. Here, attaching $e^{{\rm i}\left(q_{1}\int_{\gamma}a^{1}+q_{2}\int_{\gamma}a^{2}\right)}$ to $L_{002}\left(\gamma\right)$ does not change the original representation space of $L_{002}\left(\gamma\right)$, but only changes the characters of group elements. The fusion of two $L_{001}\left(\gamma\right)$ means the tensor product of their projective representation spaces. Thus, the fusion rule is just how to decompose the tensor product. The fusion rule takes the form of
\begin{equation}
L_{001}\left(\gamma\right)\otimes L_{001}\left(\gamma\right)=\bigoplus_{\left(q_{1}, q_{2}\right)}n_{\left(q_1, q_2\right)}L_{002}\left(\gamma\right)U_{q_1 q_2 0}\left(\gamma\right)
\end{equation}
where $q_{i}\in\left\{ 0,1,\cdots,N_{i}-1\right\} $
and $L_{002}\left(\gamma\right)U_{q_1 q_2 0}\left(\gamma\right)=L_{002}\left(\gamma\right)e^{{\rm i}\left(q_{1}\int_{\gamma}a^{1}+q_{2}\int_{\gamma}a^{2}\right)}$. Here, we use $\otimes$ and $\oplus$ to emphasize that we are dealing with the representation spaces associated with the symmetry operators.
The coefficient $n_{\left(q_1,q_2\right)}$ is given by 
\begin{align}
n_{\left(q_1, q_2\right)}= & \frac{1}{\left|\mathbb{Z}_{N_{1}}\times\mathbb{Z}_{N_{2}}\right|}\sum_{m_{1}=0}^{N_{1}-1}\sum_{m_{2}=0}^{N_{2}-1}\nonumber \\
 & \left[\chi_{\left(q_1, q_2\right)}\left(m_{1},m_{2}\right)^{*}\right]\chi_{L_{001}\left(\gamma\right)\otimes L_{001}\left(\gamma\right)}\left(m_{1},m_{2}\right)
\end{align}
where $\left(m_{1},m_{2}\right)$'s are the elements of $\mathbb{Z}_{N_{1}}\times\mathbb{Z}_{N_{2}}$
and $\chi_{a}\left(g\right)$ is the character of group element $g$
in representation $a$. The character $\chi_{L_{001}\left(\gamma\right)\otimes L_{001}\left(\gamma\right)}\left(m_{1},m_{2}\right)$
is $\text{Tr}\left[\left(U_{1}\otimes U_{1}\right)^{m_{1}}\left(U_{2}\otimes U_{2}\right)^{m_{2}}\right]=\left(\text{\text{Tr}}\left[U_{1}^{m_{1}}U_{2}^{m_{2}}\right]\right)^{2}$, where $U_{1},U_{2}$ are the clock-shift operators in the corresponding Hilbert space (see Sec.~\ref{subsec_special_type_EoM_line}), and the result is $4\cdot\delta\left(\frac{2\pi m_{1}}{N_{1}}\right)\delta\left(\frac{2\pi m_{2}}{N_{2}}\right)$.
Therefore, $\chi_{L_{001}\left(\gamma\right)\otimes L_{001}\left(\gamma\right)}\left(m_{1},m_{2}\right)=4$ when $m_{1}=0,m_{2}=0$, otherwise zero. Similarly, $\chi_{\left(q_1, q_2\right)}\left(m_{1},m_{2}\right)=e^{{\rm i}\left(\frac{2\pi q_{1}m_{1}}{N_{1}}+\frac{2\pi q_{2}m_{2}}{N_{2}}\right)}\delta\left(\frac{2\pi}{N_{1}}\cdot2m_{1}\right)\delta\left(\frac{2\pi}{N_{2}}\cdot2m_{2}\right)$.
In this way, we can calculate the values of $n_{\left(q_1, q_2\right)}$'s and find $n_{\left(0,0\right)}=1$,
$n_{\left(1,0\right)}=1$, $n_{\left(0,1\right)}=1$, $n_{\left(1,1\right)}=1$,
which gives the fusion rule
\begin{align}
L_{001}\times L_{001}= L_{002}+L_{002}U_{100}+L_{002}U_{010}+L_{002}U_{110} 
\label{eq_fusion_example_algebraic}
\end{align}
where we have neglected the supporting submanifolds of the operators for simplicity.

\subsection{A field-theoretical recipe for fusion rules\label{subsec_fusion_rule_field_theory}}

In field theory, a fusion of two topological operators means inserting
them at the same or nearby location, so that we can treat them as
a whole. The consequence is equivalent to inserting another or a linear
combination of topological operators inside the correlation function,
as dictated by the fusion rule. Let us still consider the fusion $L_{001}\left(\gamma\right)\times L_{001}\left(\gamma\right)$ in the case of $G=\mathbb{Z}_{2}\times\mathbb{Z}_{2}\times\mathbb{Z}_{4}$, as discussed in Sec.~\ref{subsec_fusion_rule_algebraic}. Inserting two $L_{001}\left(\gamma\right)$'s into a correlation
function yields
\begin{align}
 & \left\langle L_{001}\left(\gamma\right)\times L_{001}\left(\gamma\right)\mathcal{O}\right\rangle \nonumber \\
= & \left\langle 4\times e^{{\rm i}\int_{\gamma}2\cdot a^{3}+{\rm i}\Phi\left(0,0,2\right)}\times\delta\left(\int_{\gamma}a^{1}\right)\delta\left(\int_{\gamma}a^{2}\right)\mathcal{O}\right\rangle ,
\end{align}
where $\mathcal{O}$ stands for arbitrary other operators. The delta
functions inside a correlation function select the topological sectors
in the path integral. For example, in a correlation function $\left\langle \delta\left(\int_{\gamma}a^{1}\right)\mathcal{O}\right\rangle $,
since $\mathbb{Z}_{N_{1}}=\mathbb{Z}_{2}$, the values of $\int_{\gamma}a^{1}$
are $0$ or $\pi$ that label two sectors. We can expand $\delta\left(\int_{\gamma}a^{1}\right)$
as $\delta\left(\int_{\gamma}a^{1}\right)=\frac{1}{2}\left(1+e^{{\rm i}\int_{\gamma}a^{1}}\right)$
because
\begin{align}
 & \left\langle \delta\left(\int_{\gamma}a^{1}\right)\mathcal{O}\right\rangle \nonumber \\
= & \frac{1}{\mathcal{Z}}\int\mathcal{D}\left[a^{i},b^{i};\int_{\gamma}a^{1}=0\right]e^{{\rm i}S_{\text{BR}}}\mathcal{O}\nonumber \\
= & \frac{1}{\mathcal{Z}}\sum_{w\in\left\{ 0,\pi\right\} }\int\mathcal{D}\left[a^{i},b^{i};\int_{\gamma}a^{1}=w\right]e^{{\rm i}S_{\text{BR}}}\nonumber\\
 & \qquad\times\frac{1}{2}\left(1+e^{{\rm i}\int_{\gamma}a^{1}}\right)\mathcal{O},
\end{align}
where $\mathcal{D}\left[a^{i},b^{i};\int_{\gamma}a^{1}=w\right]$
denotes the configurations with $\int_{\gamma}a^{1}=w$, and $\mathcal{Z} = \int\mathcal{D}\left[a^i,b^i\right]e^{{\rm i} S_{\text{BR}}}$. The factor $4=d_{001}^{2}$ from the two auxiliary traces cancels the factor $(1/2)^2$ from expanding the two projectors, giving unit coefficients for the four fusion channels. In this way,
we find 
\begin{align}
 & \left\langle L_{001}\left(\gamma\right)\times L_{001}\left(\gamma\right)\mathcal{O}\right\rangle \nonumber \\
= & \left\langle \exp\left({\rm i}\int_{\gamma}2\cdot a^{3}+{\rm i}\Phi\left(0,0,2\right)\right)\right.\nonumber\\
&\left.\times\left(1+e^{{\rm i}\int_{\gamma}a^{1}}+e^{{\rm i}\int_{\gamma}a^{2}}+e^{{\rm i}\int_{\gamma}(a^{1}+a^{2})}\right)\mathcal{O}\right\rangle .\label{eq_L3xL3}
\end{align}
On the other hand, the unitary part of $L_{001}\left(\gamma\right)\times L_{001}\left(\gamma\right)$
is $e^{{\rm i}\int_{\gamma}2\cdot a^{3}+{\rm i}\Phi\left(0,0,2\right)}$
which is exactly the unitary part of $L_{002}\left(\gamma\right)$.
This motivates us to consider $L_{002}\left(\gamma\right)$ as the
possible fusion channels of $L_{001}\left(\gamma\right)\times L_{001}\left(\gamma\right)$.
Let us put $L_{002}\left(\gamma\right)$ into a correlation function:
\begin{align}
&\left\langle L_{002}\left(\gamma\right)\mathcal{O}\right\rangle \nonumber\\= & \left\langle e^{{\rm i}\int_{\gamma}2\cdot a^{3}+{\rm i}\Phi\left(0,0,2\right)}\delta\left(2\int_{\gamma}a^{1}\right)\delta\left(2\int_{\gamma}a^{2}\right)\mathcal{O}\right\rangle .
\end{align}
Since $\mathbb{Z}_{N_{1}}\times\mathbb{Z}_{N_{2}}=\mathbb{Z}_{2}\times\mathbb{Z}_{2}$,
the delta functions always return $1$. Thus, $\left\langle L_{002}\left(\gamma\right)\mathcal{O}\right\rangle =\left\langle e^{{\rm i}\int_{\gamma}2\cdot a^{3}+{\rm i}\Phi\left(0,0,2\right)}\mathcal{O}\right\rangle $.
Compared with Eq. (\ref{eq_L3xL3}), we find 
\begin{align}
 & \left\langle L_{001}\left(\gamma\right)\times L_{001}\left(\gamma\right)\mathcal{O}\right\rangle \nonumber \\
= & \left\langle L_{002}\left(\gamma\right)\cdot\left(1+e^{{\rm i}\int_{\gamma}a^{1}}+e^{{\rm i}\int_{\gamma}a^{2}}+e^{{\rm i}\int_{\gamma}(a^{1}+a^{2})}\right)\mathcal{O}\right\rangle .
\end{align}
Therefore, this field-theoretical analysis gives the same fusion rule
\begin{align}
L_{001}\times L_{001}= L_{002}+L_{002}U_{100}+L_{002}U_{010}+L_{002}U_{110}. 
\label{eq_fusion_example_path_integral}
\end{align}

The above discussion shows that the product of two topological operators
is equivalent to a combination of them, at the correlation function
level. This provides a physical picture of fusion that complements
the algebraic perspective. Below we introduce a systematical method
to calculate the fusion rules of non-invertible topological operators
in field theory. First, we notice that each non-invertible topological
operator has the following structure: 
\[
L_{r_{1}r_{2}r_{3}}\left(\sigma,\gamma\right) = d_{r_{1}r_{2}r_{3}}\cdot\mathbb{U}\left(r_{1},r_{2},r_{3}\right)\cdot\mathbb{P}\left(r_{1},r_{2},r_{3}\right),
\]
where $\mathbb{U}\left(r_{1},r_{2},r_{3}\right)$ is the unitary
part, i.e., $\mathbb{U}\left(r_{1},r_{2},r_{3}\right)=e^{{\rm i}\left(\int_{\sigma}r_{1}b^{1}+\int_{\sigma}r_{2}b^{2}+\int_{\gamma}r_{3}a^{3}\right)+{\rm i}\Phi\left(r_{1},r_{2},r_{3}\right)}$;
$\mathbb{P}\left(r_{1},r_{2},r_{3}\right)$ stands for the delta functions.
For the fusion of two general operators, $L_{ijk}\times L_{rst}$,
where we neglect the supporting submanifolds for simplicity, we can
deal with unitary parts and projectors separately. Since the unitary
parts appear as $c$-numbers in the path integral, the fusion of them
is a straightforward multiplication,
\begin{equation}
\left\langle \mathbb{U}\left(i,j,k\right)\times\mathbb{U}\left(r,s,t\right)\mathcal{O}\right\rangle =\left\langle \mathbb{U}\left(i+r,j+s,k+t\right)\mathcal{O}\right\rangle .
\end{equation}
The result is precisely the unitary part of the non-invertible topological
operator $L_{\left(i+r\right)\left(j+s\right)\left(k+t\right)}$. On the other hand, the fusion
of projectors yields $\mathbb{P}\left(i,j,k\right)\mathbb{P}\left(r,s,t\right)$,
which can be re-written, inside the correlation function, as a linear
combination of invertible topological operators, as in Eq. (\ref{eq_L3xL3})
for instance. Consequently, we can identify a set of invertible topological
operators $U\left(a,b,c\right)$, possibly accompanied by coefficients
$f_{abc}$, such that 
\begin{align}
 & \left\langle \mathbb{P}\left(i,j,k\right)\mathbb{P}\left(r,s,t\right)\mathcal{O}\right\rangle \nonumber\\ =&\left\langle \sum_{a,b,c}f_{abc}U\left(a,b,c\right)\mathbb{P}\left(i+r,j+s,k+t\right)\mathcal{O}\right\rangle \label{eq_delta_function_matching}
\end{align}
where $\mathbb{P}\left(i+r,j+s,k+t\right)$ is the projector of $L_{\left(i+r\right)\left(j+s\right)\left(k+t\right)}$.
Combining the contributions from the unitary parts and the projectors,
the general fusion rule takes the form
\begin{align}
& L_{ijk}\times L_{rst}\nonumber\\
=&\sum_{a,b,c}\frac{d_{ijk}d_{rst}}{d_{\left(i+r\right)\left(j+s\right)\left(k+t\right)}}f_{abc}U_{abc}L_{\left(i+r\right)\left(j+s\right)\left(k+t\right)},
\label{eq_general_fusion_rule}
\end{align}
where $d_{\left(i+r\right)\left(j+s\right)\left(k+t\right)}$ is the
quantum dimension of $L_{\left(i+r\right)\left(j+s\right)\left(k+t\right)}$. Therefore, the
essential step in determining the fusion rule is to find invertible
topological operators $U\left(a,b,c\right)$'s and corresponding $f_{abc}$'s
that satisfy Eq.~(\ref{eq_delta_function_matching}). We explain in
Appendix~\ref{Appendix_Existence_solutions_delta_function_matching}
that such $U_{abc}$'s and $f_{abc}$'s always exist. 

The $f_{abc}$ are Fourier coefficients of the projectors. To read off fusion multiplicities from Eq.~(\ref{eq_general_fusion_rule}), one must include the dimension ratio and combine equivalent output operators $U_{abc}L_{\left(i+r\right)\left(j+s\right)\left(k+t\right)}$. Two dressings give the same output when their ratio acts trivially on the holonomies selected by $\mathbb{P}\left(i+r,j+s,k+t\right)$. The resulting coefficient of each distinct output is a nonnegative integer, namely its multiplicity in the irreducible tensor-product decomposition described in Sec.~\ref{subsec_fusion_rule_algebraic}.

\section{Anomalies of (non-)invertible higher-form symmetries\label{sec_anomaly}}
    
	In this section, we analyze the anomalies of the generalized symmetries identified in previous sections. A symmetry is said to be anomalous if it cannot be gauged. For the $1$- and $2$-form symmetries studied in this paper---including both invertible and non-invertible ones---we find mixed anomalies, in the sense that certain pairs of symmetries cannot be gauged simultaneously. We first illustrate the mixed anomaly between two higher-form symmetries in the  effective field theory of $3$d $\mathbb{Z}_{N}$ toric code. We then investigate the mixed anomalies among the higher-form symmetries in the theory $S_{\text{BR}}$. We find that after gauging an invertible symmetry, an original non-invertible symmetry operator ceases to be a genuine topological operator. For pairs of non-invertible symmetries, we derive constraints that obstruct the simultaneous gauging.

	\subsection{Mixed anomalies in the $3$d $\mathbb{Z}_{N}$ topological order\label{subsec_mixed_anomaly_3D_toric_code}}
	
	In this subsection, we use the $3$d $\mathbb{Z}_{N}$ toric code as a standard example to show
	how its $1$-form and $2$-form symmetries exhibit a mixed anomaly and how the
	anomaly can be canceled by a bulk term. We start from its effective field theory
	$S_{\text{3dTC}}=\int\frac{N}{2\pi}b  da$.
	As reviewed in Sec.~\ref{subsec_review_higher-form_BF_theory}, the EoMs
	$\frac{N}{2\pi}da=0$ and $\frac{N}{2\pi}db=0$ can be formally viewed as $d\left(*\mathcal{J}_{e}\right) = 0$ and $d \left(*\mathcal{J}_{m}\right)=0$, with current analogues $\mathcal{J}_{e}=-*\frac{N}{2\pi}a$ and $\mathcal{J}_{m}=*\frac{N}{2\pi}b$. These EoMs correspond to a $1$-form symmetry with symmetry operator
	$U_{q_m}\left(\sigma\right)=e^{{\rm i}q_m \int_{\sigma}b}$
	and a $2$-form symmetry with symmetry operator
	$U_{q_e}\left(\gamma\right)=e^{{\rm i}q_{e}\int_{\gamma}a}$.
	
	We normalize the flat $\mathbb{Z}_{N}$ backgrounds $\mathcal{B}_{m}$ and $\mathcal{C}_{e}$ so that their holonomies on closed two- and three-dimensional submanifolds, respectively, lie in $\frac{2\pi}{N}\mathbb{Z}$ modulo $2\pi$. We first keep these backgrounds fixed; gauging additionally sums over their gauge-equivalence classes. The $1$-form symmetry couples to $\mathcal{B}_{m}$ via $\int_{M_4}\mathcal{B}_{m}\wedge * \mathcal{J}_{m}
	=\int_{M_4}\frac{N}{2\pi}\mathcal{B}_{m}\wedge b$.
	The total action becomes
	\begin{equation}
		S^{m}_{\text{gauged}}=\int_{M_{4}}\frac{N}{2\pi}b  da+\int_{M_4}\frac{N}{2\pi}\mathcal{B}_{m}  b.
	\end{equation}
	Under the background gauge transformation
	$\mathcal{B}_{m}\rightarrow\mathcal{B}_{m}+dV_{m}$,
	the action is invariant provided that $a\rightarrow a-V_{m}$.
	
	For the $2$-form symmetry, the coupling term is
	$\int_{M_4} \mathcal{C}_{e}\wedge* \mathcal{J}_{e}
	=\int_{M_4}\frac{N}{2\pi}\mathcal{C}_{e}\wedge a$,
	where $\mathcal{C}_{e}$ is a $3$-form background field. The total action is
	\begin{equation}
		S_{\text{gauged}}^{e}=\int_{M_{4}}\frac{N}{2\pi}b  da+\int_{M_4}\frac{N}{2\pi}\mathcal{C}_{e}  a.
        \label{eq_Zn_gauge_a}
	\end{equation}
	If $\mathcal{C}_{e}$ transforms as
	$\mathcal{C}_{e}\rightarrow\mathcal{C}_{e}+d T_{e}$,
	then $S_{\text{gauged}}^{e}$ is invariant under
	$b\rightarrow b+ T_{e}$.
	
	We now ask what happens if we attempt to gauge the $1$-form and $2$-form
	symmetries simultaneously. Including both coupling terms, we obtain
	\begin{equation}
		S_{\text{gauged}}^{m,e}=\int_{M_{4}}\frac{N}{2\pi}b  da
		+\int_{M_4}\frac{N}{2\pi}\mathcal{B}_{m}  b
		+\int_{M_4}\frac{N}{2\pi}\mathcal{C}_{e}  a.
	\label{eq_Zn_both_backgrounds}
	\end{equation}
	Under the gauge transformations $\begin{cases}
		\mathcal{B}_{m}\rightarrow & \mathcal{B}_{m}+dV_{m}\\
		a\rightarrow & a-V_{m}
	\end{cases}$ and $\begin{cases}
		\mathcal{C}_{e}\rightarrow & \mathcal{C}_{e}+dT_{e}\\
		b\rightarrow & b+ T_{e}
	\end{cases}$, the action acquires an additional term
	\begin{equation}
		\Delta S_{\text{gauged}}^{m,e}=\int_{M_{4}}\frac{N}{2\pi}\mathcal{B}_{m}T_{e}
		-\frac{N}{2\pi}\mathcal{C}_{e}V_{m}-\frac{N}{2\pi} dT_{e}V_{m}.
	\label{eq_Zn_background_variation}
	\end{equation}
	The violation of gauge invariance shows that the two symmetries cannot be gauged
	simultaneously; hence the $3$d toric code exhibits a mixed anomaly between its
	$1$-form and $2$-form symmetries.
	
	This mixed anomaly can be canceled by coupling the theory to a $(4+1)$D bulk,
	\begin{equation}
		S_{\text{bulk}}=-\int_{M_{5}}\frac{N}{2\pi}\mathcal{B}_{m} \mathcal{C}_{e}.
	\label{eq_Zn_inflow}
	\end{equation}
	Taking $\partial M_5=M_4$ and assuming that the background fields are locally flat,
	$d\mathcal{B}_{m}=0$ and $d\mathcal{C}_{e}=0$, one can verify that
	$\Delta S_{\text{bulk}} + \Delta S_{\text{gauged}}^{m,e} = 0$.

	\subsection{Mixed anomalies of higher-form symmetries in Borromean-Rings topological order\label{subsec_mixed_anomaly_aab}}
	In Sec.~\ref{sec_symmetry_operator_aab} we identified the generalized symmetries of the theory $S_{\text{BR}}$ (\ref{eq_aab_action}) and the corresponding symmetry operators. In this section, we are going to demonstrate that there exist mixed anomalies between some combinations of invertible and non-invertible higher-form symmetries, in the sense that those symmetries cannot be gauged simultaneously. Also, we will discuss linking constraints for two non-invertible higher-form symmetries in the theory $S_{{\rm BR}}$. The physical picture is that the topological excitations corresponding to these mixed-anomalous symmetry operators have nontrivial braiding statistics. For example, one of the braiding statistics can be obtained from $\left\langle U_{100}\left(\gamma\right)L_{100}\left(\sigma\right)\right\rangle_{\mathrm{norm}} =e^{-{\rm i}\frac{2\pi}{N_{1}}\text{Lk}\left(\gamma,\sigma\right)}$, where $\text{Lk}\left(\gamma,\sigma\right)$ is the linking number of $\gamma$ and $\sigma$ in the $\left(3+1\right)$D spacetime. Here, the subscript denotes normalization by the corresponding nonzero unlinked correlator.

As an example of mixed anomaly, we consider the $2$-form invertible symmetry generated by operator $U_{100}\left(\gamma\right)=e^{{\rm i}\int_{\gamma}a^{1}}$ and the $1$-form non-invertible symmetry generated by operator $L_{100}\left(\sigma\right)$. Gauging can be expressed as a sum over symmetry-defect networks with compatible fusion and junction weights. For the invertible symmetry generated by $U_{100}\left(\gamma\right)$, this gives
\begin{align}
\mathcal{Z}_{U} & =\frac{1}{\mathcal{N}}\sum_{\left[\Gamma\right]\in H_{1}\left(M_{4},\mathbb{Z}_{N_{1}}\right)}\int\mathcal{D}a^{i}\mathcal{D}b^{i}e^{{\rm i}S_{\text{BR}}}U_{100}\left(\Gamma\right)\nonumber \\
 & =\frac{\mathcal{Z}_{\mathrm{BR}}}{\mathcal{N}}\sum_{\left[\Gamma\right]\in H_{1}\left(M_{4},\mathbb{Z}_{N_{1}}\right)}\left\langle U_{100}\left(\Gamma\right)\right\rangle 
\label{eq_BR_U_network_sum}
\end{align}
where $\Gamma\equiv\sum_{i}k_{i}\gamma_{i}$ is a closed network of curves $\gamma_{i}$, $U_{100}\left(\Gamma\right)\equiv\prod_{i}U_{100}\left(\gamma_{i}\right)^{k_{i}}$, $\mathcal{Z}_{\mathrm{BR}}$ is the ungauged partition function, the brackets are normalized correlators, and $\mathcal{N}$ counts gauge redundancy. This summation must depend only on the homology class $\left[\Gamma\right]$, i.e., it is invariant under deformation of the network. For this invertible symmetry alone this holds: because the $U_{100}\left(\gamma\right)$ operators braid trivially with each other, they can freely move across each other. 

For simultaneous gauging, a candidate network sum would schematically take the form
\begin{equation}
\mathcal{Z}_{U,L}\sim\sum_{\left[\Gamma\right]\in H_{1}\left(M_{4},\mathbb{Z}_{N_{1}}\right)}\sum_{\left[\Sigma\right]}w\left(\Sigma\right)\left\langle U_{100}\left(\Gamma\right)L_{100}\left(\Sigma\right)\right\rangle  
\label{eq_candidate_joint_gauging}
\end{equation}
where $\Sigma$ denotes a network generated by $L_{100}$, including its fusion channels and junction labels, and $w(\Sigma)$ denotes the local weights required by a gauging prescription. The equivalence classes $[\Sigma]$ and the weights must be specified by consistent local network rules; they are not determined by ordinary homology alone. Equation~(\ref{eq_candidate_joint_gauging}) is a schematic candidate whose compatibility with the $U_{100}$ gauging we test below. For discussions of non-invertible gauging and its network data, see Refs.~\cite{diatlyk_gauging_2023,choi_non-invertible_2021,kaidi_kramers-wannier-like_2022,roumpedakis_higher_2023,seifnashri_gauging_2025,balasubramanian_gauging_2025,cordova_non-invertible_2024-1,antinucci_anomalies_2023,cordova_anomalies_2024}.

For the original $L_{100}$ to remain a genuine surface operator after gauging $U_{100}$, its insertion must be invariant under the $U_{100}$ gauge redundancy. Let $\mu$ be a small cycle linking one component of $\Sigma$. Although $[\Gamma+\mu]=[\Gamma]$ in the ambient spacetime, $\mu$ cannot be contracted in the complement of $\Sigma$. Contracting it through the surface gives the braiding phase
\begin{equation}
\left\langle U_{100}\left(\Gamma+\mu\right)L_{100}\left(\Sigma\right)\right\rangle =e^{-\frac{{\rm i}2\pi}{N_{1}}}\left\langle U_{100}\left(\Gamma\right)L_{100}\left(\Sigma\right)\right\rangle.
\label{eq_linked_gauge_move}
\end{equation}
For $N_1>1$, this nontrivial phase prevents the insertion of the original $L_{100}$ from being invariant under the $U_{100}$ gauging. Thus the two original defect operators cannot both be retained as genuine operators under this prescription. This crossing move tests a proposed gauge identification; ordinary topological deformations that keep the defects disjoint preserve their linking number.

The same obstruction follows directly from gauging the invertible $\mathbb{Z}_{N_1}$ 2-form symmetry. At the level of the action, we introduce $\mathcal{C}_e^1$ coupled to $a^1$, as in Eq.~(\ref{eq_Zn_gauge_a}), and integrate over it: 
\begin{equation}
S_{\text{gauged}}=S_{\text{BR}}+\int\frac{N_{1}}{2\pi}\mathcal{C}_{e}^{1}\wedge a^{1}+\frac{N_{1}}{2\pi}\varphi d\mathcal{C}_{e}^{1}
\label{eq_BR_invertible_gauging}
\end{equation}
with gauge transformation $\mathcal{C}_{e}^{1}\rightarrow\mathcal{C}_{e}^{1}+dT_{e}^{1}$ and $b^{1}\rightarrow b^{1}+T_{e}^{1}$. The Lagrange multiplier $\varphi\sim\varphi+2\pi$ enforces that the dynamical field $\mathcal{C}_{e}^{1}$ is flat with $\mathbb{Z}_{N_{1}}$ holonomies. Under $a^1\rightarrow a^1+d\chi^1$, we also take $\varphi\rightarrow\varphi-\chi^1$. In $S_{\text{gauged}}$, where the $\mathbb{Z}_{N_{1}}$ 2-form symmetry has been gauged, the gauge field $b^{1}$ shifts under the gauge transformation of $\mathcal{C}_{e}^{1}$, $b^{1}\rightarrow b^{1}+T_{e}^{1}$, so the symmetry operator $L_{100}\left(\sigma\right)$ also transforms by $e^{{\rm i}\int_{\sigma}T_{e}^{1}}$. The operator invariant under such gauge transformation is
\begin{equation}
\widetilde{V}_{100}\left(\sigma\right)=L_{100}\left(\sigma\right)e^{-{\rm i}\int_{\Omega}\mathcal{C}_{e}^{1}},\quad\partial\Omega=\sigma.
\label{eq_BR_attached_surface}
\end{equation}
The completion $\widetilde{V}_{100}$ depends on the open volume $\Omega$: changing $\Omega$ by a closed volume multiplies it by the corresponding holonomy of $\mathcal{C}_e^1$. Thus it is not a genuine surface operator supported on $\sigma$ alone. The original non-invertible operator $L_{100}$ is consequently not retained as a genuine symmetry operator after gauging $U_{100}$, in agreement with the phase in Eq.~(\ref{eq_linked_gauge_move}).

The linking phase and the attachment in Eq.~(\ref{eq_BR_attached_surface}) give two equivalent descriptions of this mixed obstruction for the original symmetry operators.

With the picture that mixed anomaly is related to nontrivial braiding statistics, we consider the linking constraints for two of the non-invertible symmetries in the theory $S_{\text{BR}}$. For instance, let us consider the symmetries generated by $L_{100}\left(\sigma\right)$ and $L_{001}\left(\gamma\right)$. The constraints associated with $L_{100}\left(\sigma\right)$ are $\delta\left(\frac{pN_{3}}{N_{123}}\int_{\sigma}b^{3}\right)$ and $\delta\left(\frac{pN_{2}}{N_{123}}\int_{S^{1}_{\tau}}a_{\tau}^{2}\right)$, and those with $L_{001}\left(\gamma\right)$ are $\delta\left(\frac{pN_{1}}{N_{123}}\int_{\gamma}a^{1}\right)$ and $\delta\left(\frac{pN_{2}}{N_{123}}\int_{\gamma}a^{2}\right)$. Suppose we insert $L_{100}\left(\sigma\right)$ and $L_{001}\left(\gamma\right)$ with linking number $l$. Integrating $b^{1}$ and $a^{3}$ yields $\frac{N_{1}}{2\pi}da^{1}=-\delta^{\perp}\left(\sigma\right)$ and $\frac{N_{3}}{2\pi}db^{3}=-\delta^{\perp}\left(\gamma\right)$. Therefore, $\int_{\sigma}b^{3}=\int-\frac{2\pi}{N_{3}}\delta^{\perp}\left(A\right)\wedge\delta^{\perp}\left(\sigma\right)=-\frac{2\pi l}{N_{3}}$, where $\partial A=\gamma$ and $\int\delta^{\perp}\left(A\right)\wedge\delta^{\perp}\left(\sigma\right)$ is exactly the linking number of $\sigma$ and $\gamma$ in spacetime. The projector has an argument $\frac{pN_{3}}{N_{123}}\int_{\sigma}b^{3}=-\frac{2\pi pl}{N_{123}}$, which is $2\pi\mathbb{Z}$ if and only if $N_{123}\mid pl$. Hence, $\left\langle L_{100}\left(\sigma\right)L_{001}\left(\gamma\right)\right\rangle =0$ unless $N_{123}\mid pl$. This is a linking selection rule. A crossing can change $l$ without changing the ambient homology classes and turn a nonzero correlator into zero. It therefore obstructs a gauging prescription that identifies these configurations while leaving the local weights unchanged. 

Before ending this section, we would like to mention that a complete analysis for gauging non-invertible symmetries requires the fusion and junction rules of the proposed networks. Identifying these data and testing the gauging procedure is an important direction for future work.

	\section{Summary and Outlook\label{sec_discussion}}
	In this work we presented a constructive route to identify generalized symmetries, including intrinsically non-invertible higher-form symmetries, in a class of $(3+1)$D twisted $BF$ topological field theories with an $a\wedge a\wedge b$ twist and gauge group $G=\prod_i \mathbb{Z}_{N_i}$. These TQFTs serve as effective continuum descriptions of Borromean-Rings topological order, i.e., three-dimensional non-Abelian topological orders supporting Borromean-Rings braiding. The starting point is the observation that an EoM of gauge field implies the topological invariance of an operator supported on a closed submanifold in the spacetime. We construct gauge invariant topological operators from the EoMs of gauge fields, which generate higher-form symmetries. Specifically, in Sec.~\ref{sec_symmetry_operator_aab}, we construct non-invertible topological line and surface operators. The non-invertibility of such operators can be understood as follows. Due to the twisted term in the theory, the integrals of some gauge fields on closed submanifolds are not gauge invariant. To recover the gauge invariance, we introduce an auxiliary system living on that submanifold so that the total system is gauge invariant. It turns out the auxiliary system transforms projectively under the gauge group, hence there is a nontrivial projective representation space associated to the operator, which makes the operator non-invertible. The correspondence among the EoMs, topological operators, and algebraic structures is summarized in Table~\ref{tab_EOM_aab}. After revealing all topological operators in the theory, we study their fusion rules and provide a field-theoretical framework to calculate the general fusion rules of topological operators in Sec.~\ref{sec_fusion_rule}. The last part of this paper, Sec.~\ref{subsec_mixed_anomaly_aab}, is for the mixed anomaly among the symmetries. We find that mixed anomalies exist for certain combinations of invertible and non-invertible symmetries, and reveal obstructions for simultaneous gauging for pairs of non-invertible symmetries.
	
	Overall, our results provide a systematic field-theoretic construction and characterization of
	invertible and non-invertible higher-form symmetries and their mixed anomalies
	in $(3+1)$D TQFTs, organized as a practical pipeline from EoMs to topological operators, fusion, and gaugeability. We expect that this construction can be extended to other families of three-dimensional non-Abelian
	topological orders (including quantum doubles of finite groups) and can serve
	as a useful interface between condensed-matter realizations and the defect/anomaly
	language natural in SymTFT. In parallel, recent progress has established
	diagrammatic representations and consistency conditions for $3$d and $4$d
	topological orders from continuum field theories~\cite{huang2025diagrammatics}.
	It would be interesting to clarify how the fusion rules derived here are
	constrained by, and in turn sharpen, these diagrammatic consistency conditions.

	Several future directions are motivated by the present study. Throughout this
	work, our analysis is primarily formulated at the level of continuum field
	theory. An important next step is to formulate and test generalized symmetries
	in concrete lattice models that realize three-dimensional topological orders.
	Natural candidates include three-dimensional quantum double models, Hamiltonian
	lattice realizations of $(3+1)$D Dijkgraaf--Witten theories, and three-dimensional
	generalizations of string-net constructions~\cite{Kitaev2003faulttolerant,wan_twisted_2015,Huxford_membrane,string_net_levin_wen_2005}.
	Recently, the authors of this work investigated microscopic constructions of
	excitations, fusion processes, and shrinking operations in non-Abelian
	topological orders using the three-dimensional quantum double
	model~\cite{huang2025bridge}. Building on this line of work, it would be
	valuable to explicitly construct invertible and non-invertible higher-form
	symmetry operators on three-dimensional lattices and to systematically study
	their properties, including anomalies and onsitability, directly at the
	microscopic
	level~\cite{choi_non-invertible_2025,kobayashi2024generalized,feng_higher-form_2025,tantivasadakarn_sequential_2025}.
	Such constructions would provide a concrete bridge between continuum effective
	field theories and lattice Hamiltonians.
	
	Another natural direction is to combine the results of this work to the SymTFT framework. The next step is to explore how the generalized symmetries of the $\left(3+1\right)$D TQFT $S_{\text{BR}}$ impose constraints for its $\left(2+1\right)$D boundary theory. A systematic analysis of boundary theories from this
	perspective will lead to a unified understanding of bulk-boundary correspondence in higher dimensions. In addition, it is intriguing to explore
	whether the $1$-form and $2$-form symmetries identified in this work organize
	into a nontrivial higher-group structure~\cite{barkeshli_higher-group_2024,cordova_exploring_2019,Liu_2025_higher_matter}. Clarifying such higher-group structures, both in continuum field theories and in lattice realizations, could further illuminate the internal symmetry organization of three-dimensional topological orders.
	
	Beyond symmetry structure, a promising set of directions concerns the interplay
	between generalized symmetries and quantum-information properties. In
	particular, it would be interesting to study how generalized symmetries
	constrain the entanglement structure of quantum states. For example, one may
	investigate the circuit complexity required to prepare quantum states that
	differ by their generalized symmetry content, potentially revealing intrinsic
	notions of complexity protected by higher-form or non-invertible symmetries.
	Sequential quantum circuits for topological code-state preparation~\cite{Hu_SEESQC_2025} provide a concrete setting for exploring this direction.
	Furthermore, extending the theory of anyon condensation to $(3+1)$D systems
	remains an important open problem. Such a generalization is expected to be
	closely related to phase transitions between $(3+1)$D topological phases
	characterized by distinct generalized symmetry structures.
	
	Finally, it is important to explore how generalized symmetries constrain the
	behavior of quantum systems at finite temperature or under decoherence, where
	the system is in a mixed state rather than a pure state. Subdimensional reduced states of stabilizer models also exhibit strong and weak symmetries determined by bulk stabilizers~\cite{Li_2025_Subdim_EE}, providing a complementary route to mixed-state symmetry and holography. Recent work has shown
	that three-dimensional systems below a nonzero temperature can still exhibit
	topological order due to anomalous $2$-form symmetries associated with emergent
	fermionic excitations~\cite{zhou2025finite}. Since $BF$ theories with a
	$b\wedge b$ twist describe a class of three-dimensional topological orders with
	fermionic excitations~\cite{PhysRevB.99.235137, bti2,
		Kapustin:2014gua, zhang_continuum_2023}, they provide a natural arena to further investigate the
	role of anomalous generalized symmetries in realizing mixed-state topological
	order. Understanding these phenomena within a unified continuum
	field-theoretical framework may shed new light on the robustness of topological
	phases beyond the zero-temperature, closed-system setting.

	\acknowledgments 
	This work was in part supported by National Natural Science Foundation of China (NSFC)  under Grants No. 12474149 and No. 12274250. %Research Center for Magnetoelectric Physics of Guangdong Province under Grant No. 2024B0303390001, and Guangdong Provincial Key Laboratory of Magnetoelectric Physics and Devices under Grant No. 2022B1212010008.
	
	\appendix
	
	\onecolumngrid

\section{Generalized symmetries in lattice models\label{appendix_symm_lattice}}

	In Sec.~\ref{sec_review_generalized_symmetry} we explain the concepts of generalized symmetries in the language of continuum field theory. As a comparison, in this appendix, we use two representative lattice model examples to illustrate the higher-form symmetry (the toric code models in $2$d and $3$d) and the non-invertible symmetry (the $1$d critical Ising model).
	\subsection{Higher-form symmetries in $2$d and $3$d toric code models}
	\label{appendix_higher_form_symm_toric_code}
	
	The Hamiltonian of the toric code model on a square lattice is given by
	\begin{equation}
		H_{\text{TC}}
		=-\sum_{v}A_{v}-\sum_{p}B_{p}
		=-\sum_{v}\prod_{\partial l\ni v}X_{l}
		-\sum_{p}\prod_{l\in\partial p}Z_{l},
	\end{equation}
	where qubits reside on the links of the lattice.
	The ground state satisfies $A_{v}=B_{p}=1$ for all vertices $v$ and plaquettes $p$.
	An excitation with $A_{v}=-1$ ($B_{p}=-1$) is referred to as an $e$ excitation
	(an $m$ excitation).
	
	There exist two operators that commute with $H_{\text{TC}}$,
	\begin{equation}
		W_{e}\left(L\right)=\prod_{l\in L}Z_{l},
		\qquad
		W_{m}\left(L^{*}\right)=\prod_{l\perp L^{*}}X_{l},
	\end{equation}
	where $L$ and $L^{*}$ are closed loops on the original lattice and its dual,
	respectively.
	Each operator generates a $\mathbb{Z}_{2}$ $1$-form symmetry, denoted as
	$\mathbb{Z}_{2,e}^{(1)}$ and $\mathbb{Z}_{2,m}^{(1)}$, respectively.
	
	We can identify operators charged under the
	$\mathbb{Z}_{2,e}^{(1)}$ and $\mathbb{Z}_{2,m}^{(1)}$ symmetry transformations,
	\begin{equation}
		O_{x}=\prod_{l\perp c^{*}}X_{l},
		\qquad
		O_{z}=\prod_{l\in c}Z_{l},
	\end{equation}
	where $c^{*}$ ($c$) is a line on the dual (original) lattice.
	The lines $c^{*}$ and $c$ need not be closed.
	The actions of the symmetry operators on these charged operators are
	\begin{equation}
		W_{e}\left(L\right)O_{x}W_{e}^{\dagger}\left(L\right)
		=\left(-1\right)^{{\rm Int}\left(L,c^{*}\right)}O_{x},
	\end{equation}
	\begin{equation}
		W_{m}\left(L^{*}\right)O_{z}W_{m}^{\dagger}\left(L^{*}\right)
		=\left(-1\right)^{{\rm Int}\left(L^{*},c\right)}O_{z},
	\end{equation}
	where ${\rm Int}\left(L,c^{*}\right)$ denotes the intersection number of $L$ and $c^{*}$.
	When $c^*$ ($c$) is an open line, the $X$ ($Z$) string creates a pair of $m$ ($e$) excitations, consistently with the plaquette ($B_p$) and vertex ($A_v$) terms in the Hamiltonian. Accordingly, $W_e(L)$ measures the number of $m$ excitations modulo $2$ enclosed by $L$, while $W_m(L^*)$ measures the number of $e$ excitations modulo $2$ enclosed by $L^*$.
	
	In addition, the two symmetry operators satisfy
	\begin{equation}
		W_{e}\left(L\right)W_{m}\left(L^{*}\right)
		=\left(-1\right)^{{\rm Int}\left(L,L^{*}\right)}
		W_{m}\left(L^{*}\right)W_{e}\left(L\right).
	\end{equation}
	This relation implies that $W_{e}\left(L\right)$ is charged under the symmetry
	generated by $W_{m}\left(L^{*}\right)$, and vice versa. These operators generate a $\mathbb{Z}_{2,e}^{(1)}\times\mathbb{Z}_{2,m}^{(1)}$
	$1$-form symmetry, consistent with the lattice description.
	
	We next turn to the $3$d toric code model to illustrate $1$-form and $2$-form
	symmetries.
	Consider a three-dimensional cubic lattice with qubits on the links.
	The Hamiltonian is
	\begin{equation}
		H_{\text{3dTC}}
		=-\sum_{v}A_{v}-\sum_{p}B_{p}
		=-\sum_{v}\prod_{l\ni v}X_{l}
		-\sum_{p}\prod_{l\in\partial p}Z_{l}.
	\end{equation}
	A particle excitation at vertex $v$ corresponds to $A_{v}=-1$.
	A pair of particle excitations can be created at the endpoints of an open string
	$s$ by applying
	$W_{e}\left(s\right)=\prod_{l\in s}Z_{l}$ to the ground state.
	A loop excitation is created by applying a surface operator
	$W_{m}\left(A^{*}\right)=\prod_{l\perp A^{*}}X_{l}$,
	where $A^{*}$ is an open surface on the dual lattice.
	
	Each $X_l$ operator flips the four plaquette terms sharing the link $l$ on the cubic lattice. These four plaquettes correspond to the four dual links bounding the dual plaquette pierced by $l$. In the product $W_m(A^*)$, the flips on interior dual links cancel in pairs, while those along $\partial A^*$ remain. Thus the plaquettes dual to $\partial A^*$ have $B_p=-1$, producing a loop excitation along the boundary of $A^*$.
	
	Analogous to the $2$d case, the $3$d toric code model hosts two higher-form
	symmetries.
	One is generated by
	\begin{equation}
		W_{e}\left(L\right)=\prod_{l\in L}Z_{l},
	\end{equation}
	where $L$ is a closed curve on the cubic lattice.
	This operator acts on a codimension-$2$ submanifold and generates a $2$-form
	symmetry.
	Its effect can be interpreted as moving a particle excitation along $L$.
	The other symmetry is generated by
	\begin{equation}
		W_{m}\left(M^{*}\right)=\prod_{l\perp M^{*}}X_{l},
	\end{equation}
	where $M^{*}$ is a closed surface on the dual lattice.
	This operator acts on a codimension-$1$ submanifold and generates a $1$-form
	symmetry, corresponding to moving a loop excitation along $M^{*}$.
	The two symmetry operators satisfy
	\begin{equation}
		W_{e}\left(L\right)W_{m}\left(M^{*}\right)
		=\left(-1\right)^{{\rm Int}\left(L,M^{*}\right)}
		W_{m}\left(M^{*}\right)W_{e}\left(L\right),
	\end{equation}
	where ${\rm Int}\left(L,M^{*}\right)$ denotes the intersection number of $L$ and $M^{*}$. These operators generate a $\mathbb{Z}_{2}$ $2$-form and a $\mathbb{Z}_{2}$
	$1$-form symmetry, respectively.
	
	An equivalent description of the $3$d toric code places qubits on the plaquettes
	of the cubic lattice.
	The Hamiltonian is
	\begin{equation}
		H_{\text{3dTC}}^{\prime}
		=-\sum_{c}A_{c}-\sum_{l}B_{l}
		=-\sum_{c}\prod_{p\in\partial c}X_{p}
		-\sum_{l}\prod_{p\ni l}Z_{p}.
	\end{equation}
	The two descriptions are related by the identity
	\begin{equation}
		S=\int_{M_{4}}\frac{2}{2\pi}bda
		=\int_{M_{4}}\frac{2}{2\pi}adb
		+\text{(exterior derivative)}.
	\end{equation}
	In this field-theoretical description, the higher-form symmetries are encoded in
	the flatness conditions $da=0$ and $db=0$, which ensure the topological invariance of the Wilson operators.
	The coefficient $\frac{1}{\pi}$ reflects the underlying $\mathbb{Z}_{2}$ structure.
	
	\subsection{Non-invertible symmetry in $(1+1)$D transverse-field Ising model}
	\label{appendix_review_non-invertible}
	
	In Appendix~\ref{appendix_higher_form_symm_toric_code}, we reviewed generalized
	symmetries whose operators form a group structure and therefore admit inverses.
	A complementary generalization of symmetry relaxes the requirement of
	invertibility.
	Symmetry operators of this type do not possess inverses and are referred to as
	non-invertible symmetries.
	
	The $(1+1)$D transverse-field Ising model at its critical point provides a
	canonical example of a non-invertible symmetry
	\cite{bhardwaj2018finite,chang_topological_2019,shao_whats_2024}.
	The Hamiltonian is
	\begin{equation}
		H=-\sum_{i=1}^{N}Z_{i}Z_{i+1}-g\sum_{i=1}^{N}X_{i},
	\end{equation}
	with periodic boundary conditions $X_{N+1}=X_{1}$ and $Z_{N+1}=Z_{1}$.
	This model has a $\mathbb{Z}_{2}$ global symmetry generated by
	$\mathsf{P}=\prod_{i=1}^{N}X_{i}$, which satisfies
	$\left[\mathsf{P},H\right]=0$ and $\mathsf{P}^{2}=\mathbb{I}$.
	
	At the critical point $g=1$, the Hamiltonian is invariant under the
	Kramers--Wannier transformation,
	\begin{equation}
		X_{i}\rightarrow Z_{i}Z_{i+1},
		\qquad
		Z_{i}Z_{i+1}\rightarrow X_{i+1},
		\qquad i=1,\ldots,N.
	\end{equation}
	Although one might attempt to realize this transformation using a unitary
	operator, such an operator does not exist.
	
	If such a unitary operator $U$ existed, satisfying $UHU^{-1}=H$ and
	$U X_{i} U^{-1} = Z_i Z_{i+1}$, then it would imply
	\begin{equation}
		U \mathsf{P} U^{-1}
		=U\left(\prod_{i=1}^{N}X_{i}\right)U^{-1}
		=\prod_{i=1}^{N}Z_{i}Z_{i+1}
		=1,
	\end{equation}
	which would force $\mathsf{P}=1$ and lead to a contradiction.
	Therefore, the Kramers--Wannier transformation cannot be implemented by a unitary
	operator.
	
	Instead, the Kramers--Wannier transformation is realized by a non-invertible
	operator $\mathsf{D}$ acting on the Hilbert space,
	\begin{equation}
		\mathsf{D}=U_{\rm KW}\frac{1+\mathsf{P}}{2},
	\end{equation}
	where
\begin{align}
U_{\rm KW}
&=e^{-\frac{{\rm i}2\pi N}{8}}
\left(\prod_{i=1}^{N-1}\frac{1+{\rm i}X_i}{\sqrt{2}}\frac{1+{\rm i}Z_iZ_{i+1}}{\sqrt{2}}\right)
\frac{1+{\rm i}X_N}{\sqrt{2}}
\end{align}
is unitary. The factors in the ordered product are written from left to right with increasing $i$, so that the final $X_N$ factor acts first on a state. The factor $(1+\mathsf{P})/2$ projects onto the $\mathsf{P}=+1$ sector.
	The operator $\mathsf{D}$ acts on local operators as
	\begin{equation}
		\mathsf{D} X_{i} = Z_{i} Z_{i+1}\mathsf{D},
		\qquad
		\mathsf{D} Z_{i} Z_{i+1} = X_{i+1} \mathsf{D},
		\qquad i = 1,\cdots, N,
	\end{equation}
	with $X_{N+1} = X_{1}$ and $Z_{N+1}= Z_{1}$, which reproduces the
	Kramers--Wannier transformation.
	
	For comparison, the action of $U_{\rm KW}$ is
	$U_{\rm KW} X_{i} U_{\rm KW}^{-1} =  Z_{i}Z_{i+1}$ and
	$U_{\rm KW} Z_{i} Z_{i+1} U_{\rm KW}^{-1} =  X_{i+1}$ for $i=1,\cdots,N-1$,
	while for the boundary terms one finds
	$U_{\rm KW} X_{N} U_{\rm KW}^{-1} =  \mathsf{P} Z_{N}Z_{1}$ and
	$U_{\rm KW} Z_{N} Z_{1} U_{\rm KW}^{-1} =  \mathsf{P} X_{1}$.
	Hence $U_{\rm KW}$ does not commute with the Hamiltonian and is not a symmetry.
	By contrast, $\mathsf{D}$ commutes with $H$ at $g=1$,
	\begin{equation}
		\mathsf{D}H = H\mathsf{D},
		\qquad
		\text{or}\qquad
		\left[\mathsf{D},H\right]=0,
		\qquad
		\text{when } g=1.
	\end{equation}
	
	On the other hand, because $\mathsf{D}$ contains a projector, its action
	annihilates all $\mathbb{Z}_{2}$-odd states.
	Therefore, there is no operator $\mathsf{O}$ such that
	$\mathsf{O}\mathsf{D}\left|\psi\right\rangle=\left|\psi\right\rangle$.
	In this sense, $\mathsf{D}$ is non-invertible, and the associated symmetry is
	referred to as a non-invertible symmetry.
	
	The key ingredient underlying this non-invertible symmetry is the invariance of
	the critical theory under the Kramers--Wannier transformation, i.e., under gauging.
	It has also been found that certain $(3+1)$D theories exhibit a related invariance
	under gauging: one may gauge only part of the system and thereby create a
	topological defect interface separating the original and gauged regions.
	Such topological defects are non-invertible
	\cite{tachikawa_gauging_2020, choi_non-invertible_2021, koide_noninvertible_2023}.
	In the present example, the symmetry operator $\mathsf{D}$ is a product of a
	unitary operator $U_{\rm KW}$ and a projector, where the latter is responsible
	for the non-invertibility.

    \section{Construction of topological operators from type-II EoMs\label{Appendix_Calculation_partition_function_auxiliary_system}}
The theory we are studying is $S_{\text{BR}}=\int\sum_{i=1}^{3}\frac{N_{i}}{2\pi}b^{i}da^{i}+qa^{1}a^{2}b^{3}$,
where $a^{i}$ and $b^{i}$ are $1$-form and $2$-form gauge fields,
respectively. It is a $\left(3+1\right)$D twisted $BF$ gauge theory
with gauge group $G=\prod_{i=1}^{3}\mathbb{Z}_{N_{i}}$. The coefficient is
$q=\frac{pN_{1}N_{2}N_{3}}{\left(2\pi\right)^{2}N_{123}}$, where
$N_{123}$ is the greatest common divisor of $N_{1},N_{2},N_{3}$,
and $p\in\mathbb{Z}_{N_{123}}$. Unless stated otherwise, we assume $p\neq0$
in $\mathbb{Z}_{N_{123}}$. For $p=0$, the theory reduces to the
untwisted $BF$ theory. This theory is invariant under the following
gauge transformations: $a^{1}\rightarrow a^{1}+d\chi^{1}$, $a^{2}\rightarrow a^{2}+d\chi^{2}$,
$b^{3}\rightarrow b^{3}+dV^{3}$, and 
\begin{align}
a^{3}\rightarrow & a^{3}+d\chi^{3}-\frac{2\pi q}{N_{3}}\left(\chi^{1}a^{2}+\frac{1}{2}\chi^{1}d\chi^{2}\right)+\frac{2\pi q}{N_{3}}\left(\chi^{2}a^{1}+\frac{1}{2}\chi^{2}d\chi^{1}\right),\nonumber \\
b^{1}\rightarrow & b^{1}+dV^{1}-\frac{2\pi q}{N_{1}}\left(\chi^{2}b^{3}-a^{2}V^{3}+\chi^{2}dV^{3}\right),\nonumber \\
b^{2}\rightarrow & b^{2}+dV^{2}+\frac{2\pi q}{N_{2}}\left(\chi^{1}b^{3}-a^{1}V^{3}+\chi^{1}dV^{3}\right),
\end{align}
where $\chi^{i}$ and $V^{i}$ are $0$-form and $1$-form gauge parameters
with $\int d\chi^{i}\in2\pi\mathbb{Z}$ and $\int dV^{i}\in2\pi\mathbb{Z}$,
respectively. As mentioned in the main text, we call the equations
of motion (EoMs) for $a^{3},b^{1},b^{2}$, obtained by varying $b^3,a^1,a^2$, respectively, type-II EoMs:
$\frac{N_{3}}{2\pi}da^{3}+qa^{1}a^{2}=0$, $\frac{N_{1}}{2\pi}db^{1}+qa^{2}b^{3}=0$,
$\frac{N_{2}}{2\pi}db^{2}-qa^{1}b^{3}=0$.

\subsection{Topological lines \label{Appendix_line_construction}}

In this appendix, we derive Eq.~(\ref{eq_L001}). We parameterize $\gamma$ by $\tau\in\left[\tau_{0},\tau_{f}\right)$
and treat $\tau$ as time.

\emph{Quantization.} Canonical quantization of $S_{\text{1d aux}}$
in Eq.~(\ref{eq_S_1d_aux}) gives $\left[\phi^{1},\phi^{2}\right]=\frac{{\rm i}2\pi}{K_{3}}$
and $H_{\text{1d aux}}=\frac{K_{3}}{2\pi}\left(-\phi^{1}a^{2}+\phi^{2}a^{1}\right)$. From now on, $\phi^{1}$ and $\phi^{2}$ are quantum operators that are not dependent on time,  rather than the field variables in the Lagrangian where time dependence is explicit. After canonical quantization, the two gauge fields $a^{1}$ and $a^2$ as background fields are still time-dependent field variables along $\gamma$. 

With only $a^{1}$ present, the evolution along $\gamma$ leads to  $e^{-\frac{{\rm i}K_{3}}{2\pi}\phi^{2}\int_{\gamma}a^{1}}=U_{1}^{k_{1}}$, where  $U_1$ and $\int_\gamma a^1$ are given below Eq.~(\ref{eqexplain}). Thus $U_{1}e^{i\phi^{1}}U_{1}^{-1}=e^{i\phi^{1}}e^{-i\frac{2\pi}{N_{1}}}$ and we have $U_{1}^{N_{1}}=\mathds{1}$. Likewise, $a^{2}$ alone leads to $e^{\frac{{\rm i}K_{3}}{2\pi}\phi^{1}\int_{\gamma}a^{2}}=U_{2}^{k_{2}}$, where  $U_2$ and $\int_\gamma a^2$ are given below Eq.~(\ref{eqexplain}). 
Thus $U_{2}e^{i\phi^{2}}U_{2}^{-1}=e^{i\phi^{2}}e^{-i\frac{2\pi}{N_{2}}}$, and we have $U_{2}^{N_{2}}=\mathds{1}$.   Applying the commutator between $\phi^{1}$ and $\phi^{2}$ yields the algebra (\ref{eq_line_projective_algebra}), $U_{1} U_{2} = e^{-\frac{{\rm i}2 \pi p}{N_{123}}}U_{2}U_{1}$, 
 which can be realized on the $d$-dimensional space $\mathcal{H}$ with basis (\ref{eq_line_clock_basis}).
For the chosen irreducible auxiliary sector, the partition function can be rewritten as
\begin{equation}
\mathcal{Z}_{\text{1d aux}}\left(a^{1},a^{2}\right)=\text{Tr}_{\mathcal H}\left[\alpha(g)^{-1}\mathcal{U}\right],\label{eq_appendix_Z_1d_aux}
\end{equation}
with $\mathcal{U}=\mathcal{T}e^{-{\rm i}\int_{\tau_{0}}^{\tau_{f}}d\tau H_{\text{1d aux}}\left(\tau\right)}$
the $\tau$-ordered evolution operator along $\gamma$. The factor $\alpha(g)^{-1}$ implements the ordered holonomy convention defined below; it equals one at trivial holonomy, preserving $\mathrm{Tr}_{\mathcal H}\mathbf{1}=d$.

\emph{Time-ordered exponential.} Since $\left[H_{\text{1d aux}}\left(\tau\right),H_{\text{1d aux}}\left(\mu\right)\right]$
is a $c$-number, the Magnus expansion terminates at second order
(this holds for all auxiliary systems considered in this paper),
giving
\begin{equation}
\mathcal{U}=e^{\frac{{\rm i}K_{3}}{2\pi}\left(\phi^{1}\int_{\gamma}a^{2}-\phi^{2}\int_{\gamma}a^{1}\right)}\,e^{{\rm i}\Phi},\label{eq_appendix_line_Magnus}
\end{equation}
where $e^{{\rm i}\Phi}$ is a pure phase with $\Phi=\frac{K_{3}}{4\pi}\int_{\gamma}\left(\widetilde{\phi}^{1}a^{2}-\widetilde{\phi}^{2}a^{1}\right)$
and $\widetilde{\phi}^{i}\left(\tau\right)=\int_{\tau_{0}}^{\tau}a^{i}\left(\mu\right)d\mu$. By the Baker-Campbell-Hausdorff (BCH) formula, the operator
part is
\begin{equation}
e^{\frac{{\rm i}K_{3}}{2\pi}\left(\phi^{1}\int_{\gamma}a^{2}-\phi^{2}\int_{\gamma}a^{1}\right)}=U_{2}^{k_{2}}U_{1}^{k_{1}}\,e^{-\frac{{\rm i}\pi pk_{1}k_{2}}{N_{123}}}.\label{eq_appendix_line_BCH_phase}
\end{equation}
The extra phase $\alpha\left(g\right)=e^{-\frac{{\rm i}\pi pk_{1}k_{2}}{N_{123}}}$,
$g=\left(k_{1},k_{2}\right)\in\mathbb{Z}_{N_{1}}\times\mathbb{Z}_{N_{2}}$,
only relabels the projective representation $\rho\left(g\right)=U_{2}^{k_{2}}U_{1}^{k_{1}}$
to $\rho'\left(g\right)=\alpha\left(g\right)\rho\left(g\right)$. Their $2$-cocycles satisfy
$\omega(g,h)=e^{-\frac{{\rm i}2\pi p k_1l_2}{N_{123}}}$ and
$\omega'(g,h)=\alpha(g)\alpha(h)\alpha(g+h)^{-1}\omega(g,h)$,
where $h=(l_1,l_2)$ and $g+h$ is reduced modulo $(N_1,N_2)$ using the chosen integer representatives. Thus they differ by the coboundary $\delta\alpha$ and define the same class in $H^2(\mathbb{Z}_{N_1}\times\mathbb{Z}_{N_2},U(1))$. We use the ordered representative $\rho(g)$ for the auxiliary background coupling; the compensating rephasing $\alpha(g)^{-1}$ in Eq.~(\ref{eq_appendix_Z_1d_aux}) makes this convention explicit. On the support of the projectors, $\alpha(g)$ can be $-1$ as well as $1$, so this rephasing is a convention for the coupling, not a consequence of the projectors. The same convention is used for the surface and composite defects below.

\emph{Trace.} In the clock-shift basis,
\begin{equation}
\text{Tr}_{\mathcal{H}}\left[U_{2}^{k_{2}}U_{1}^{k_{1}}\right]=\sum_{n=0}^{d-1}e^{-\frac{{\rm i}2\pi pk_{1}n}{N_{123}}}\left\langle n|n+k_{2}\right\rangle ,
\end{equation}
which equals $d$ if $k_{1}=k_{2}=0\mod d$ and vanishes otherwise.
Since $k_{i}=0\mod d$ is equivalent to $\frac{pk_{i}}{N_{123}}\in\mathbb{Z}$,
i.e., $\frac{pN_{i}}{N_{123}}\int_{\gamma}a^{i}\in2\pi\mathbb{Z}$,
we obtain
\begin{align}
\mathcal{Z}_{\text{1d aux}}\left(a^{1},a^{2}\right)= & \,d\cdot e^{{\rm i}\Phi}\nonumber \\
 & \times\delta\left(\frac{pN_{1}}{N_{123}}\int_{\gamma}a^{1}\right)\delta\left(\frac{pN_{2}}{N_{123}}\int_{\gamma}a^{2}\right),
\end{align}
which, combined with $e^{{\rm i}\int_{\gamma}a^{3}}$, gives Eq.~(\ref{eq_L001}).

\subsection{Topological surfaces\label{Appendix_surface_construction}}

Here we derive Eq.~(\ref{eq_L100}) for a general charge $r_{1}\in\mathbb{Z}_{N_{1}}$,
i.e., $L_{r_{1}00}\left(\sigma\right)=e^{{\rm i}r_{1}\int_{\sigma}b^{1}}\mathcal{Z}_{\text{2d aux}}\left(r_{1};a^{2},b^{3}\right)$,
whose auxiliary action $S_{\text{2d aux}}\left(r_{1};a^{2},b^{3}\right)$
is Eq.~(\ref{eq_S_2d_aux}) with $K_{1}\rightarrow r_{1}K_{1}$. For $r_1=0$ we set $L_{000}=1$; the following quantization applies to $r_1\ne0$. The
steps parallel Appendix~\ref{Appendix_line_construction}.

\emph{Dimensional reduction.} On $\sigma=S_{\eta}^{1}\times S_{\tau}^{1}$
with orientation $d\tau\wedge d\eta$, we write $\beta^{3}=\beta_{\tau}^{3}d\tau+\beta_{\eta}^{3}d\eta$,
$a^{2}=a_{\tau}^{2}d\tau+a_{\eta}^{2}d\eta$, and $b^{3}=b_{\tau\eta}^{3}d\tau\wedge d\eta$.
Integrating by parts along $S_{\eta}^{1}$,
\begin{equation}
S_{\text{2d aux}}\left(r_{1};a^{2},b^{3}\right)=\frac{r_{1}K_{1}}{2\pi}\int d\tau d\eta\left[-\phi^{2}\partial_{\tau}\beta_{\eta}^{3}+\beta_{\tau}^{3}\left(a_{\eta}^{2}-\partial_{\eta}\phi^{2}\right)-\beta_{\eta}^{3}a_{\tau}^{2}+\phi^{2}b_{\tau\eta}^{3}\right],
\end{equation}
where $\beta_{\tau}^{3}$ is a Lagrange multiplier imposing $\partial_{\eta}\phi^{2}=a_{\eta}^{2}$.
For $a_{\eta}^{2}=0$ and $a_{\tau}^{2}$ independent of $\eta$, we
have $\phi^{2}=\phi^{2}\left(\tau\right)$, and with $B^{3}\left(\tau\right)=\int_{S_{\eta}^{1}}\beta_{\eta}^{3}d\eta$
and $\Theta^{3}\left(\tau\right)=\int_{S_{\eta}^{1}}b_{\tau\eta}^{3}d\eta$,
\begin{equation}
S_{\text{2d aux}}=\frac{r_{1}K_{1}}{2\pi}\int d\tau\left(-\phi^{2}\partial_{\tau}B^{3}-B^{3}a_{\tau}^{2}+\phi^{2}\Theta^{3}\right).
\end{equation}

\emph{Quantization.} This gives $\left[\phi^{2},B^{3}\right]=\frac{{\rm i}2\pi}{r_{1}K_{1}}$
and $H_{\text{2d aux}}=\frac{r_{1}K_{1}}{2\pi}\left(B^{3}a_{\tau}^{2}-\phi^{2}\Theta^{3}\right)$.
With $\int_{S_{\tau}^{1}}a_{\tau}^{2}=\frac{2\pi k_{2}}{N_{2}}$ and
$\int_{S_{\tau}^{1}}\Theta^{3}=\int_{\sigma}b^{3}=\frac{2\pi k_{3}}{N_{3}}$,
$a^{2}$ alone gives the evolution $U_{2}^{k_{2}}$ with $U_{2}=e^{-\frac{{\rm i}r_{1}K_{1}}{N_{2}}B^{3}}$
and $U_{2}\phi^{2}U_{2}^{-1}=\phi^{2}-\frac{2\pi}{N_{2}}$, while
$b^{3}$ alone gives $U_{3}^{k_{3}}$ with $U_{3}=e^{\frac{{\rm i}r_{1}K_{1}}{N_{3}}\phi^{2}}$
and $U_{3}B^{3}U_{3}^{-1}=B^{3}-\frac{2\pi}{N_{3}}$. The compactness of $\phi^{2}$ and $\beta^{3}$ requires $U_{2}^{N_{2}}=U_{3}^{N_{3}}=\mathds{1}$. The commutator gives the algebra $U_{2}U_{3}=e^{-\frac{{\rm i}2\pi r_{1}p}{N_{123}}}U_{3}U_{2}$,
which reduces to Eq.~(\ref{eq_surface_projective_algebra}) for $r_{1}=1$.
The irreducible representation space $\mathcal{H}$ has dimension $d=\frac{N_{123}}{\gcd\left(r_{1}p,N_{123}\right)}$
and basis $U_{2}|n\rangle=e^{-\frac{{\rm i}2\pi r_{1}pn}{N_{123}}}|n\rangle$,
$U_{3}|n\rangle=|n+1\rangle$.

\emph{Trace.} As in Appendix~\ref{Appendix_line_construction}, $\mathcal{Z}_{\text{2d aux}}=\text{Tr}_{\mathcal{H}}\left[e^{-\frac{{\rm i}\pi r_1p k_2k_3}{N_{123}}}\mathcal{U}\right]$
with $\mathcal{U}=\mathcal{T}e^{-{\rm i}\int_{\tau_{0}}^{\tau_{f}}d\tau H_{\text{2d aux}}}$,
and the Magnus expansion gives
\begin{equation}
\mathcal{U}=e^{-\frac{{\rm i}r_{1}K_{1}}{2\pi}\left(B^{3}\int_{S_{\tau}^{1}}a_{\tau}^{2}-\phi^{2}\int_{\sigma}b^{3}\right)}\,e^{{\rm i}\Phi},
\end{equation}
where $\Phi=\frac{r_{1}K_{1}}{4\pi}\left(\int_{\sigma}b^{3}\widetilde{\phi}^{2}-\int_{S_{\tau}^{1}}a_{\tau}^{2}\widetilde{\beta}^{3}\right)$,
with $\widetilde{\phi}^{2}\left(\tau\right)=\int_{\tau_{0}}^{\tau}a_{\tau}^{2}\left(\mu\right)d\mu$
and $\widetilde{\beta}^{3}\left(\tau\right)=\int_{\tau_{0}}^{\tau}\Theta^{3}\left(\mu\right)d\mu$.
By the BCH formula, the operator part equals $U_{2}^{k_{2}}U_{3}^{k_{3}}e^{\frac{{\rm i}\pi r_{1}pk_{2}k_{3}}{N_{123}}}$,
whose scalar phase is canceled by the prefactor in the trace, implementing the holonomy convention specified below Eq.~(\ref{eq_appendix_line_BCH_phase}).
The trace $\text{Tr}_{\mathcal{H}}\left[U_{2}^{k_{2}}U_{3}^{k_{3}}\right]$
equals $d$ if $k_{2}=k_{3}=0\mod d$ and vanishes otherwise, so
\begin{align}
L_{r_{1}00}\left(\sigma\right)= & \,d\cdot e^{{\rm i}\int_{\sigma}r_{1}b^{1}+{\rm i}\Phi}\nonumber \\
 & \times\delta\left(\frac{r_{1}pN_{3}}{N_{123}}\int_{\sigma}b^{3}\right)\delta\left(\frac{r_{1}pN_{2}}{N_{123}}\int_{S_{\tau}^{1}}a_{\tau}^{2}\right),\label{eq_appendix_L_r1_00}
\end{align}
which reduces to Eq.~(\ref{eq_L100}) for $r_{1}=1$ and is non-invertible
for $d>1$.

\emph{Surface operator from EoM of $b^{2}$.} The construction for $L_{0r_{2}0}\left(\sigma\right)=e^{{\rm i}r_{2}\int_{\sigma}b^{2}}\mathcal{Z}_{\text{2d aux}}^{\prime}\left(r_{2};a^{1},b^{3}\right)$
is identical, with the auxiliary action
\begin{equation}
S_{\text{2d aux}}^{\prime}=-\int_{\sigma}\frac{r_{2}K_{2}}{2\pi}\left(-\phi^{1}d\beta^{3}+\beta^{3}a^{1}+\phi^{1}b^{3}\right).
\end{equation}
The result is
\begin{align}
L_{0r_{2}0}\left(\sigma\right)= & \,d\cdot e^{{\rm i}\int_{\sigma}r_{2}b^{2}-\frac{{\rm i}r_{2}K_{2}}{4\pi}\left(\int_{\sigma}b^{3}\widetilde{\phi}^{1}-\int_{S_{\tau}^{1}}a_{\tau}^{1}\widetilde{\beta}^{3}\right)}\nonumber \\
 & \times\delta\left(\frac{r_{2}pN_{3}}{N_{123}}\int_{\sigma}b^{3}\right)\delta\left(\frac{r_{2}pN_{1}}{N_{123}}\int_{S_{\tau}^{1}}a_{\tau}^{1}\right),
\end{align}
with $d=\frac{N_{123}}{\gcd\left(r_{2}p,N_{123}\right)}$ and $\widetilde{\phi}^{1}\left(\tau\right)=\int_{\tau_{0}}^{\tau}a_{\tau}^{1}\left(\mu\right)d\mu$.
For $r_{2}=1$ it reduces to Eq.~(\ref{eq_L010}).

\subsection{A general family of topological operators\label{Appendix_general_construction}}

Here we derive Eq.~(\ref{eq_general_L}) for
\begin{align}
L_{r_{1}r_{2}r_{3}}\left(\sigma,\gamma\right)= & \,e^{{\rm i}\int_{\sigma}\left(r_{1}b^{1}+r_{2}b^{2}\right)+{\rm i}\int_{\gamma}r_{3}a^{3}}\times\mathcal{Z}_{\text{aux}}\left(r_{1},r_{2},r_{3};a^{1},a^{2},b^{3}\right),\label{eq_appendix_general_op_expression}
\end{align}
where $\sigma=S_{\eta}^{1}\times S_{\tau}^{1}$ and $\gamma$ is the
non-contractible circle along $S_{\tau}^{1}$. The auxiliary action
combines those of $L_{r_{1}00}$, $L_{0r_{2}0}$, and $L_{00r_{3}}$:
\begin{align}
S_{\text{aux}}= & \int_{\sigma}\frac{r_{1}K_{1}}{2\pi}\left(-\phi^{2}d\beta^{3}+\beta^{3}a^{2}+\phi^{2}b^{3}\right)\nonumber \\
 & -\int_{\sigma}\frac{r_{2}K_{2}}{2\pi}\left(-\phi^{1}d\beta^{3}+\beta^{3}a^{1}+\phi^{1}b^{3}\right)\nonumber \\
 & +\int_{\gamma}\frac{r_{3}K_{3}}{2\pi}\left(-\phi^{1}d\phi^{2}+\phi^{1}a^{2}-\phi^{2}a^{1}\right).
\end{align}
For $r_1=r_2=r_3=0$, the operator is the identity. The following derivation assumes a nonzero charge vector. When only $r_3$ is nonzero it reduces to Appendix~\ref{Appendix_line_construction}; when only $r_1$ or $r_2$ is nonzero it reduces to Appendix~\ref{Appendix_surface_construction}.

\emph{Dimensional reduction.} As in Appendix~\ref{Appendix_surface_construction},
integrating by parts along $S_{\eta}^{1}$ turns $\beta_{\tau}^{3}$
into a Lagrange multiplier imposing $r_{1}K_{1}\left(a_{\eta}^{2}-\partial_{\eta}\phi^{2}\right)=r_{2}K_{2}\left(a_{\eta}^{1}-\partial_{\eta}\phi^{1}\right)$.
For $\phi^{1,2}$ depending only on $\tau$ and $a_{\tau}^{1,2}$ independent
of $\eta$, and with $B^{3}$ and $\Theta^{3}$ defined in Appendix~\ref{Appendix_surface_construction},
\begin{align}
S_{\text{aux}}= & \int d\tau\frac{r_{1}K_{1}}{2\pi}\left(-\phi^{2}\partial_{\tau}B^{3}-B^{3}a_{\tau}^{2}+\phi^{2}\Theta^{3}\right)\nonumber \\
 & -\int d\tau\frac{r_{2}K_{2}}{2\pi}\left(-\phi^{1}\partial_{\tau}B^{3}-B^{3}a_{\tau}^{1}+\phi^{1}\Theta^{3}\right)\nonumber \\
 & +\int d\tau\frac{r_{3}K_{3}}{2\pi}\left(-\phi^{1}\partial_{\tau}\phi^{2}+\phi^{1}a_{\tau}^{2}-\phi^{2}a_{\tau}^{1}\right).
\end{align}

\emph{Quantization.} The kinetic terms couple three variables $\left(\phi^{1},\phi^{2},B^{3}\right)$
through a rank-$2$ form. Let $g=\gcd\left(r_{1}K_{1},r_{2}K_{2},r_{3}K_{3}\right)$
and $\vec{n}=\left(r_{1}K_{1},r_{2}K_{2},r_{3}K_{3}\right)/g$, and
choose integer vectors $\vec{f},\vec{h}$ with $\vec{f}\times\vec{h}=\vec{n}$.\footnote{With $x,y\in\mathbb{Z}$ such that $n_{1}x+n_{2}y=\gcd\left(n_{1},n_{2}\right)$
(B\'ezout), one can take $\vec{f}=\left(-n_{2},n_{1},0\right)/\gcd\left(n_{1},n_{2}\right)$
and $\vec{h}=\left(-n_{3}x,-n_{3}y,\gcd\left(n_{1},n_{2}\right)\right)$. If $n_1=n_2=0$, then $n_3=1$ and one can instead take $\vec f=(1,0,0)$ and $\vec h=(0,1,0)$.} In terms of $\theta^{1}=f_{1}\phi^{1}+f_{2}\phi^{2}+f_{3}B^{3}$
and $\theta^{2}=h_{1}\phi^{1}+h_{2}\phi^{2}+h_{3}B^{3}$, one finds
\begin{align}
S_{\text{aux}}=  -\frac{g}{4\pi}\int_{S_{\tau}^{1}}\left(\theta^{1}d\theta^{2}-\theta^{2}d\theta^{1}\right)+\frac{g}{2\pi}\int_{S_{\tau}^{1}}\left(\theta^{1}\,\vec{h}\cdot\vec{A}-\theta^{2}\,\vec{f}\cdot\vec{A}\right),
\end{align}
where $\vec{A}=\left(a_{\tau}^{1},a_{\tau}^{2},\Theta^{3}\right)$.
Hence $\left[\theta^{1},\theta^{2}\right]=\frac{{\rm i}2\pi}{g}$ and
$H_{\text{aux}}=-\frac{g}{2\pi}\left(\theta^{1}\,\vec{h}\cdot\vec{A}-\theta^{2}\,\vec{f}\cdot\vec{A}\right)$.
The operators $U\left(V\right)=e^{\frac{{\rm i}g}{2\pi}\left(v_{2}\theta^{1}-v_{1}\theta^{2}\right)}$ with $V=\left(v_1, v_2\right)$
shift $\left(\theta^{1},\theta^{2}\right)$ by $\left(-v_{1},-v_{2}\right)$
and obey
\begin{equation}
U\left(V\right)U\left(V^{\prime}\right)=e^{-\frac{{\rm i}g}{4\pi}V\wedge V^{\prime}}U\left(V+V^{\prime}\right),\label{eq_appendix_general_U_commutator}
\end{equation}
with $V\wedge V^{\prime}=v_{1}v_{2}^{\prime}-v_{2}v_{1}^{\prime}$.
For holonomies $\int_{S_{\tau}^{1}}a^{1,2}=\frac{2\pi k_{1,2}}{N_{1,2}}$
and $\int_{\sigma}b^{3}=\frac{2\pi k_{3}}{N_{3}}$,
\begin{equation}
e^{-{\rm i}\int_{S_{\tau}^{1}}H_{\text{aux}}}=U\left(\Delta\left(\vec{k}\right)\right),\label{eq_appendix_general_exp_-iH}
\end{equation}
where $\Delta\left(\vec{k}\right)=\sum_{i=1}^{3}k_{i}\Delta_{i}$
and $\Delta_{i}=\frac{2\pi}{N_{i}}\left(f_{i},h_{i}\right)$. The operator
$U_{i}\equiv U\left(\Delta_{i}\right)$ shifts $\phi^{1}$, $\phi^{2}$,
$B^{3}$ by $-\frac{2\pi}{N_{i}}$ for $i=1,2,3$, respectively, and
\begin{align}
U\left(\Delta\left(\vec{k}\right)\right)U\left(\Delta\left(\vec{k}^{\prime}\right)\right)
 =e^{-{\rm i}2\pi\vec{s}\cdot\left(\vec{k}\times\vec{k}^{\prime}\right)}U\left(\Delta\left(\vec{k}^{\prime}\right)\right)U\left(\Delta\left(\vec{k}\right)\right),\label{eq_appendix_U_k_commutator}
\end{align}
with $\vec{s}=\frac{p}{N_{123}}\left(r_{1},r_{2},r_{3}\right)$. In
particular, $U_{1}U_{2}=e^{-\frac{{\rm i}2\pi r_{3}p}{N_{123}}}U_{2}U_{1}$,
$U_{2}U_{3}=e^{-\frac{{\rm i}2\pi r_{1}p}{N_{123}}}U_{3}U_{2}$, and
$U_{3}U_{1}=e^{-\frac{{\rm i}2\pi r_{2}p}{N_{123}}}U_{1}U_{3}$, reproducing
the algebra of Sec.~\ref{subsec_special_type_EoM_line} and \ref{subsec_special_type_EoM_surface}, and Appendices~\ref{Appendix_line_construction} and \ref{Appendix_surface_construction}.

\emph{Hilbert space.} Let $\hat{r}=\left(r_{1},r_{2},r_{3}\right)/\gcd\left(r_{1},r_{2},r_{3}\right)$
and choose integer vectors $\vec{u},\vec{t}$ with $\det\left(\hat{r},\vec{u},\vec{t}\right)=1$.\footnote{Write $\hat{r}=\left(a,b,c\right)$, $g^{\prime}=\gcd\left(b,c\right)$,
and choose $\beta,\gamma,x,y\in\mathbb{Z}$ with $\beta b+\gamma c=g^{\prime}$
and $xa+yg^{\prime}=1$ [possible since $\gcd\left(a,g^{\prime}\right)=1$].
Then $\vec{u}=\left(0,\gamma,-\beta\right)$ and $\vec{t}=\left(-y,xb/g^{\prime},xc/g^{\prime}\right)$
satisfy $\hat{r}\cdot\left(\vec{u}\times\vec{t}\right)=1$ when $g^{\prime}>0$. If $b=c=0$, then $a=1$ and one can take $\vec u=(0,1,0)$ and $\vec t=(0,0,1)$.} Any $\vec{k}$ decomposes as
\begin{equation}
\vec{k}=c_{0}\hat{r}+c_{1}\vec{u}+c_{2}\vec{t},\quad c_{i}\in\mathbb{Z},\label{eq_appendix_k_decomposition}
\end{equation}
with $c_{1}=\left(\hat{r}\times\vec{k}\right)\cdot\vec{t}$ and $c_{2}=-\left(\hat{r}\times\vec{k}\right)\cdot\vec{u}$.
Since $\vec{s}\parallel\hat{r}$, $U\left(\Delta\left(\hat{r}\right)\right)$
is central, while $Z\equiv U\left(\Delta\left(\vec{u}\right)\right)$
and $X\equiv U\left(\Delta\left(\vec{t}\right)\right)$ satisfy $ZX=\omega XZ$
with $\omega=e^{-\frac{{\rm i}2\pi p\gcd\left(r_{1},r_{2},r_{3}\right)}{N_{123}}}$.
The order of $\omega$ is 
\begin{equation}
d=\frac{N_{123}}{\gcd\left(N_{123},r_{1}p,r_{2}p,r_{3}p\right)},
\end{equation}
which is the dimension of the irreducible representation $\mathcal{H}$,
with clock-shift basis $Z|n\rangle=\omega^{n}|n\rangle$, $X|n\rangle=|n+1\rangle$.

\emph{Trace.} As in Appendices~\ref{Appendix_line_construction} and
\ref{Appendix_surface_construction}, $\mathcal{Z}_{\text{aux}}=\text{Tr}_{\mathcal{H}}\left[\widetilde{\xi}(\vec k)^{-1}\mathcal{U}\right]$, with the phase convention specified below,
and the Magnus expansion gives
\begin{equation}
\mathcal{U}=U\left(\Delta\left(\vec{k}\right)\right)e^{{\rm i}\Phi\left(r_{1},r_{2},r_{3}\right)},\label{eq_appendix_general_time_evolution_op}
\end{equation}
with $\Phi\left(r_{1},r_{2},r_{3}\right)$ given in Eq.~(\ref{eq_general_Phi}).
Using Eqs.~(\ref{eq_appendix_general_U_commutator}) and (\ref{eq_appendix_k_decomposition}),
$U\left(\Delta\left(\vec{k}\right)\right)=\widetilde{\xi}\left(\vec{k}\right)Z^{c_{1}}X^{c_{2}}$,
where $\widetilde{\xi}\left(\vec{k}\right)=e^{\frac{{\rm i}\pi c_{1}c_{2}p\gcd\left(r_{1},r_{2},r_{3}\right)}{N_{123}}}U\left(\Delta\left(\hat{r}\right)\right)^{c_{0}}$
is a $c$-number. Then
\begin{equation}
\text{Tr}_{\mathcal{H}}\left[Z^{c_{1}}X^{c_{2}}\right]=\sum_{n=0}^{d-1}\omega^{c_{1}\left(n+c_{2}\right)}\left\langle n|n+c_{2}\right\rangle ,\label{eq_appendix_trace_U_k}
\end{equation}
which equals $d$ if $c_{1},c_{2}=0\mod d$ and vanishes otherwise.
From Eq.~(\ref{eq_appendix_k_decomposition}) and $\hat{r}\times\vec{k}=c_{1}\hat{r}\times\vec{u}+c_{2}\hat{r}\times\vec{t}$,
the condition $c_{1},c_{2}=0\mod d$ is equivalent to $\hat{r}\times\vec{k}=0\mod d$.
Since $\vec{s}=\frac{p\gcd\left(r_{1},r_{2},r_{3}\right)}{N_{123}}\hat{r}$,
this is in turn equivalent to
\begin{equation}
\left(\vec{s}\times\vec{k}\right)_{l}=\epsilon_{lij}\frac{r_{i}pk_{j}}{N_{123}}\in\mathbb{Z},\quad l=1,2,3,\label{eq_appendix_k_constraint_equivalent}
\end{equation}
which, with $k_{1,2}=\frac{N_{1,2}}{2\pi}\int_{\gamma}a^{1,2}$ and
$k_{3}=\frac{N_{3}}{2\pi}\int_{\sigma}b^{3}$, are exactly the three
delta functions in $\mathbb{P}\left(r_{1},r_{2},r_{3}\right)$, Eq.~(\ref{eq_general_P}).
We select the central character $U(\Delta(\hat r))=1$ for this elementary defect and compensate the remaining BCH phase by $\widetilde{\xi}(\vec k)^{-1}$ in the auxiliary background coupling, as in Appendix~\ref{Appendix_line_construction}. This fixes the phase convention while retaining the ordinary trace normalization. Therefore,
$\mathcal{Z}_{\text{aux}}=d\cdot e^{{\rm i}\Phi\left(r_{1},r_{2},r_{3}\right)}\cdot\mathbb{P}\left(r_{1},r_{2},r_{3}\right)$,
which gives Eq.~(\ref{eq_general_L}).

\section{Existence of solutions for Eq. (\ref{eq_delta_function_matching})\label{Appendix_Existence_solutions_delta_function_matching}}

We show that there exist $U\left(a,b,c\right)$'s and $f_{abc}$'s
satisfying
\begin{equation}
\left\langle \mathbb{P}\left(i,j,k\right)\mathbb{P}\left(r,s,t\right)\mathcal{O}\right\rangle =\left\langle \sum_{a,b,c}f_{abc}U\left(a,b,c\right)\mathbb{P}\left(i+r,j+s,k+t\right)\mathcal{O}\right\rangle .
\end{equation}
Below we write the two operators as $\mathbb{P}\left(r_{1},r_{2},r_{3}\right)$
and $\mathbb{P}\left(e_{1},e_{2},e_{3}\right)$. Each $\mathbb{P}\left(r_{1},r_{2},r_{3}\right)$
consists of three delta functions. We first consider one of them,
$\delta\left(\frac{r_{1}pN_{2}}{N_{123}}\int_{\gamma}a^{2}-\frac{r_{2}pN_{1}}{N_{123}}\int_{\gamma}a^{1}\right)$.
With $\int_{\gamma}a^{i}=\frac{2\pi k_{i}}{N_{i}}$, this delta function
requires $\frac{2\pi p}{N_{123}}\left(r_{1}k_{2}-r_{2}k_{1}\right)\in2\pi\mathbb{Z}$,
i.e., $r_{1}k_{2}-r_{2}k_{1}=0\mod{\mathsf{d}}$, with $\mathsf{d}=\frac{N_{123}}{\gcd\left(N_{123},p\right)}$.
So, we have
\begin{align}
\delta\left(\frac{r_{1}pN_{2}}{N_{123}}\int_{\gamma}a^{2}-\frac{r_{2}pN_{1}}{N_{123}}\int_{\gamma}a^{1}\right) & =\frac{1}{\mathsf{d}}\sum_{l=0}^{\mathsf{d}-1}e^{{\rm i}l\cdot\left(\frac{r_{1}pN_{2}}{N_{123}}\int_{\gamma}a^{2}-\frac{r_{2}pN_{1}}{N_{123}}\int_{\gamma}a^{1}\right)}.\label{eq_appC_delta_Fourier}
\end{align}
Therefore, we can write the product of two delta functions, which
is a part of $\mathbb{P}\left(r_{1},r_{2},r_{3}\right)\mathbb{P}\left(e_{1},e_{2},e_{3}\right)$,
as
\begin{align}
\mathbb{P}_{\overline{r_{3}}}\mathbb{P}_{\overline{e_{3}}}\equiv & \delta\left(\frac{r_{1}pN_{2}}{N_{123}}\int_{\gamma}a^{2}-\frac{r_{2}pN_{1}}{N_{123}}\int_{\gamma}a^{1}\right)\delta\left(\frac{e_{1}pN_{2}}{N_{123}}\int_{\gamma}a^{2}-\frac{e_{2}pN_{1}}{N_{123}}\int_{\gamma}a^{1}\right)\nonumber \\
= & \frac{1}{\mathsf{d}^{2}}\sum_{l_{r},l_{e}=0}^{\mathsf{d}-1}e^{{\rm i}\left[\frac{\left(l_{r}r_{1}+l_{e}e_{1}\right)pN_{2}}{N_{123}}\int_{\gamma}a^{2}-\frac{\left(l_{r}r_{2}+l_{e}e_{2}\right)pN_{1}}{N_{123}}\int_{\gamma}a^{1}\right]}.\label{eq_appC_two_delta}
\end{align}
On the other hand,
\begin{align}
\mathbb{P}_{\overline{r_{3}+e_{3}}}\equiv & \delta\left(\frac{\left(r_{1}+e_{1}\right)pN_{2}}{N_{123}}\int_{\gamma}a^{2}-\frac{\left(r_{2}+e_{2}\right)pN_{1}}{N_{123}}\int_{\gamma}a^{1}\right)\nonumber \\
= & \frac{1}{\mathsf{d}}\sum_{l_{3}=0}^{\mathsf{d}-1}e^{{\rm i}l_{3}\left[\frac{\left(r_{1}+e_{1}\right)pN_{2}}{N_{123}}\int_{\gamma}a^{2}-\frac{\left(r_{2}+e_{2}\right)pN_{1}}{N_{123}}\int_{\gamma}a^{1}\right]},\label{eq_appC_sum_delta}
\end{align}
which is one of the three delta functions in $\mathbb{P}\left(r_{1}+e_{1},r_{2}+e_{2},r_{3}+e_{3}\right)$.

To relate Eqs.~(\ref{eq_appC_two_delta}) and (\ref{eq_appC_sum_delta}),
we define
\begin{equation}
A_{\vec{r}}\equiv\frac{r_{1}pN_{2}}{N_{123}}\int_{\gamma}a^{2}-\frac{r_{2}pN_{1}}{N_{123}}\int_{\gamma}a^{1},
\end{equation}
so that $\mathbb{P}_{\overline{r_{3}}}=\delta\left(A_{\vec{r}}\right)$,
$\mathbb{P}_{\overline{e_{3}}}=\delta\left(A_{\vec{e}}\right)$, and
$\mathbb{P}_{\overline{r_{3}+e_{3}}}=\delta\left(A_{\vec{r}+\vec{e}}\right)$.
Two properties of $A_{\vec{r}}$ are needed. First, it is linear
in $\vec{r}$, $A_{\vec{r}+\vec{e}}=A_{\vec{r}}+A_{\vec{e}}$. Second,
$A_{\vec{r}}=\frac{2\pi p}{N_{123}}\left(r_{1}k_{2}-r_{2}k_{1}\right)$,
so that $\mathsf{d}A_{\vec{r}}=\frac{2\pi p}{\gcd\left(N_{123},p\right)}\left(r_{1}k_{2}-r_{2}k_{1}\right)\in2\pi\mathbb{Z}$,
i.e., $e^{{\rm i}\mathsf{d}A_{\vec{r}}}=1$. The exponent in Eq.~(\ref{eq_appC_two_delta})
can then be rewritten as
\begin{equation}
l_{r}A_{\vec{r}}+l_{e}A_{\vec{e}}=\left(l_{r}-l_{e}\right)A_{\vec{r}}+l_{e}A_{\vec{r}+\vec{e}}.
\end{equation}
We change the summation variables to $l_{1}\equiv l_{r}-l_{e}\mod{\mathsf{d}}$
and $l_{3}\equiv l_{e}$. The map $\left(l_{r},l_{e}\right)\mapsto\left(l_{1},l_{3}\right)$
is a bijection of $\mathbb{Z}_{\mathsf{d}}\times\mathbb{Z}_{\mathsf{d}}$,
and reducing $l_{r}-l_{e}$ modulo $\mathsf{d}$ does not change the
summand because $e^{{\rm i}\mathsf{d}A_{\vec{r}}}=1$. Therefore,
\begin{equation}
\sum_{l_{r},l_{e}=0}^{\mathsf{d}-1}e^{{\rm i}\left(l_{r}A_{\vec{r}}+l_{e}A_{\vec{e}}\right)}=\sum_{l_{1}=0}^{\mathsf{d}-1}e^{{\rm i}l_{1}A_{\vec{r}}}\times\sum_{l_{3}=0}^{\mathsf{d}-1}e^{{\rm i}l_{3}A_{\vec{r}+\vec{e}}}.\label{eq_appC_resummation}
\end{equation}
Note that the same index $l_{1}$ multiplies both $\int_{\gamma}a^{2}$
and $\int_{\gamma}a^{1}$, and likewise for $l_{3}$; the phases depending
on $a^{1}$ and on $a^{2}$ cannot be resummed separately. Substituting
Eq.~(\ref{eq_appC_resummation}) into Eq.~(\ref{eq_appC_two_delta})
and using Eq.~(\ref{eq_appC_sum_delta}), we obtain
\begin{equation}
\mathbb{P}_{\overline{r_{3}}}\mathbb{P}_{\overline{e_{3}}}=\frac{1}{\mathsf{d}}\sum_{l=0}^{\mathsf{d}-1}e^{{\rm i}l\left(\frac{r_{1}pN_{2}}{N_{123}}\int_{\gamma}a^{2}-\frac{r_{2}pN_{1}}{N_{123}}\int_{\gamma}a^{1}\right)}\,\mathbb{P}_{\overline{r_{3}+e_{3}}}\label{eq_single_delta_match_a1a2_result}
\end{equation}
with $\mathsf{d}=\frac{N_{123}}{\gcd\left(N_{123},p\right)}$. Each
term in the sum is an invertible type-I operator $e^{{\rm i}m_{1}\int_{\gamma}a^{1}+{\rm i}m_{2}\int_{\gamma}a^{2}}$
with integer charges $m_{1}=-\frac{lr_{2}pN_{1}}{N_{123}}$ and $m_{2}=\frac{lr_{1}pN_{2}}{N_{123}}$.
The prefactor in Eq.~(\ref{eq_single_delta_match_a1a2_result}) is
just the expansion (\ref{eq_appC_delta_Fourier}) of $\mathbb{P}_{\overline{r_{3}}}$,
so Eq.~(\ref{eq_single_delta_match_a1a2_result}) is equivalent to
$\mathbb{P}_{\overline{r_{3}}}\mathbb{P}_{\overline{e_{3}}}=\mathbb{P}_{\overline{r_{3}}}\mathbb{P}_{\overline{r_{3}+e_{3}}}$.
This is expected: once $A_{\vec{r}}\in2\pi\mathbb{Z}$, the condition
$A_{\vec{e}}\in2\pi\mathbb{Z}$ is equivalent to $A_{\vec{r}+\vec{e}}\in2\pi\mathbb{Z}$.

Let us generalize the above discussion to $\mathbb{P}\left(r_{1},r_{2},r_{3}\right)\mathbb{P}\left(e_{1},e_{2},e_{3}\right)$
and $\mathbb{P}\left(r_{1}+e_{1},r_{2}+e_{2},r_{3}+e_{3}\right)$.
We define the arguments of the other two delta functions,
\begin{align}
B_{\vec{r}} & \equiv\frac{r_{1}pN_{3}}{N_{123}}\int_{\sigma}b^{3}-\frac{r_{3}pN_{1}}{N_{123}}\int_{\gamma}a^{1},\\
C_{\vec{r}} & \equiv\frac{r_{2}pN_{3}}{N_{123}}\int_{\sigma}b^{3}-\frac{r_{3}pN_{2}}{N_{123}}\int_{\gamma}a^{2},
\end{align}
so that $\mathbb{P}\left(r_{1},r_{2},r_{3}\right)=\mathbb{P}_{\overline{r_{1}}}\mathbb{P}_{\overline{r_{2}}}\mathbb{P}_{\overline{r_{3}}}=\delta\left(C_{\vec{r}}\right)\delta\left(B_{\vec{r}}\right)\delta\left(A_{\vec{r}}\right)$.
With $\int_{\sigma}b^{3}=\frac{2\pi k_{3}}{N_{3}}$, $B_{\vec{r}}$
and $C_{\vec{r}}$ are also linear in $\vec{r}$ and satisfy $e^{{\rm i}\mathsf{d}B_{\vec{r}}}=e^{{\rm i}\mathsf{d}C_{\vec{r}}}=1$.
Repeating the argument above for each of the three pairs gives $\mathbb{P}_{\overline{r_{i}}}\mathbb{P}_{\overline{e_{i}}}=\mathbb{P}_{\overline{r_{i}}}\mathbb{P}_{\overline{r_{i}+e_{i}}}$
($i=1,2,3$), hence
\begin{equation}
\mathbb{P}\left(r_{1},r_{2},r_{3}\right)\mathbb{P}\left(e_{1},e_{2},e_{3}\right)=\mathbb{P}\left(r_{1},r_{2},r_{3}\right)\mathbb{P}\left(r_{1}+e_{1},r_{2}+e_{2},r_{3}+e_{3}\right).
\end{equation}
Here $r_{i}+e_{i}$ may be taken modulo $N_{i}$, since $\mathbb{P}$
is invariant under $r_{i}\rightarrow r_{i}+N_{i}$ (every argument
then shifts by $\frac{2\pi pN_{i}k_{j}}{N_{123}}\in2\pi\mathbb{Z}$).
Expanding each delta function in $\mathbb{P}\left(r_{1},r_{2},r_{3}\right)$
as in Eq.~(\ref{eq_appC_delta_Fourier}), with Fourier indices $n_{1},n_{2},n_{3}$
for $\delta\left(C_{\vec{r}}\right),\delta\left(B_{\vec{r}}\right),\delta\left(A_{\vec{r}}\right)$,
respectively, we find
\begin{equation}
\mathbb{P}\left(r_{1},r_{2},r_{3}\right)\mathbb{P}\left(e_{1},e_{2},e_{3}\right)=\frac{1}{\mathsf{d}^{3}}\sum_{n_{1},n_{2},n_{3}=0}^{\mathsf{d}-1}U\left(m_{1},m_{2},m_{3}\right)\,\mathbb{P}\left(r_{1}+e_{1},r_{2}+e_{2},r_{3}+e_{3}\right),\label{eq_appC_full_result}
\end{equation}
where $U\left(m_{1},m_{2},m_{3}\right)=e^{{\rm i}m_{1}\int_{\gamma}a^{1}+{\rm i}m_{2}\int_{\gamma}a^{2}+{\rm i}m_{3}\int_{\sigma}b^{3}}$
is an invertible type-I operator with integer charges
\begin{align}
m_{1} & =-\frac{pN_{1}}{N_{123}}\left(n_{2}r_{3}+n_{3}r_{2}\right),\nonumber \\
m_{2} & =\frac{pN_{2}}{N_{123}}\left(n_{3}r_{1}-n_{1}r_{3}\right),\nonumber \\
m_{3} & =\frac{pN_{3}}{N_{123}}\left(n_{1}r_{2}+n_{2}r_{1}\right).
\end{align}
Therefore, we can identify a set of invertible topological operators
$U\left(a,b,c\right)$, accompanied by coefficients $f_{abc}$, such
that
\begin{align}
\left\langle \mathbb{P}\left(r_{1},r_{2},r_{3}\right)\mathbb{P}\left(e_{1},e_{2},e_{3}\right)\mathcal{O}\right\rangle =\left\langle \sum_{m_{1},m_{2},m_{3}}f_{m_{1}m_{2}m_{3}}U\left(m_{1},m_{2},m_{3}\right)\mathbb{P}\left(r_{1}+e_{1},r_{2}+e_{2},r_{3}+e_{3}\right)\mathcal{O}\right\rangle ,
\end{align}
where $f_{m_{1}m_{2}m_{3}}$ equals $\frac{1}{\mathsf{d}^{3}}$ times
the number of $\left(n_{1},n_{2},n_{3}\right)\in\mathbb{Z}_{\mathsf{d}}^{3}$
that give the charges $\left(m_{1},m_{2},m_{3}\right)$ modulo $\left(N_{1},N_{2},N_{3}\right)$.
These coefficients are not unique, because some $U\left(m_{1},m_{2},m_{3}\right)$
act trivially on $\mathbb{P}\left(r_{1}+e_{1},r_{2}+e_{2},r_{3}+e_{3}\right)$; such terms must be grouped into distinct output operators. Including the dimension ratio in Eq.~(\ref{eq_general_fusion_rule}) then gives the fusion multiplicities, as explained in Sec.~\ref{subsec_fusion_rule_field_theory}.

\twocolumngrid

%\bibliography{SymTFT2}


\end{document}